\documentclass[twoside,twocolumn,french,english,8.5pt]{article}
\usepackage[super,sort&compress,comma]{natbib} 
\usepackage{mhchem}
\usepackage{times,mathptmx}
\usepackage{sectsty}
\usepackage{balance} 

\usepackage{graphicx} 
\usepackage{lastpage}
\usepackage[format=plain,justification=raggedright,singlelinecheck=false,font=small,labelfont=bf,labelsep=space]{caption} 
\usepackage{fancyhdr}
\usepackage{color}

\usepackage[T1]{fontenc}
\usepackage[latin9]{inputenc}
\usepackage{float}
\usepackage{units}
\usepackage{amsmath}
\usepackage{amssymb}
\usepackage{esint}
\PassOptionsToPackage{normalem}{ulem}
\usepackage{ulem}

\makeatletter
\@ifundefined{showcaptionsetup}{}{%
 \PassOptionsToPackage{caption=false}{subfig}}
\usepackage{subfig}
\makeatother

\usepackage{babel}
\addto\extrasfrench{%
   
   \providecommand{\fg}{\ifdim\lastskip>\z@\unskip\fi~\frqq}
}

\begin{document}

\def\NOTE#1{{\textcolor{blue}{ [#1]}}}  
\def\REPLACE#1{{\textcolor{red}{ #1}}}  

\thispagestyle{plain}
\fancypagestyle{plain}{
\renewcommand{\headrulewidth}{1pt}}
\renewcommand{\thefootnote}{\fnsymbol{footnote}}
\renewcommand\footnoterule{\vspace*{1pt}%
\hrule width 3.4in height 0.4pt \vspace*{5pt}} 
\setcounter{secnumdepth}{5}

\makeatletter 
\def\subsubsection{\@startsection{subsubsection}{3}{10pt}{-1.25ex plus -1ex minus -.1ex}{0ex plus 0ex}{\normalsize\bf}} 
\def\paragraph{\@startsection{paragraph}{4}{10pt}{-1.25ex plus -1ex minus -.1ex}{0ex plus 0ex}{\normalsize\textit}} 
\renewcommand\@makefntext[1]%
{\noindent\makebox[0pt][r]{\@thefnmark\,}#1}
\makeatother 
\renewcommand{\figurename}{\small{Fig.}~}
\sectionfont{\large}
\subsectionfont{\normalsize} 

\fancyfoot{}
\fancyfoot[RO]{\footnotesize{\sffamily{1--\pageref{LastPage} ~\textbar  \hspace{2pt}\thepage}}}
\fancyfoot[LE]{\footnotesize{\sffamily{\thepage~\textbar\hspace{3.45cm} 1--\pageref{LastPage}}}}
\fancyhead{}
\renewcommand{\headrulewidth}{1pt} 
\renewcommand{\footrulewidth}{1pt}
\setlength{\arrayrulewidth}{1pt}
\setlength{\columnsep}{6.5mm}
\setlength\bibsep{1pt}

\twocolumn[
  \begin{@twocolumnfalse}
\noindent\LARGE{\textbf{Universal and non-universal features in coarse-grained models of flow in disordered solids}}
\vspace{0.6cm}

\noindent\large{\textbf{Alexandre Nicolas,\textit{$^{a,b}$} Kirsten Martens,\textit{$^{a,b}$} Lyd\'eric Bocquet,\textit{$^{c}$} and
Jean-Louis Barrat\textit{$^{a,b,d}$}}}\vspace{0.5cm}

\noindent\textit{\small{\textbf{Received Xth XXXXXXXXXX 20XX, Accepted Xth XXXXXXXXX 20XX\newline
First published on the web Xth XXXXXXXXXX 200X}}}

\noindent \textbf{\small{DOI: 10.1039/b000000x}}
\vspace{0.6cm}

\noindent \normalsize{We study the two-dimensional (2D) shear flow of amorphous solids within
variants of an elastoplastic model, paying particular attention to
spatial correlations and time fluctuations of, e.g., local stresses. The model is based on
the local alternation between an elastic regime and plastic events
during which the local stress is redistributed. The importance of
a fully tensorial description of the stress and of the inclusion of
(coarse-grained) convection in the model is investigated; scalar and
tensorial models yield very similar results, while convection enhances
fluctuations and breaks the spurious symmetry between the flow and
velocity gradient directions, for instance when shear localisation
is observed. Besides, correlation lengths measured with diverse protocols
are discussed. One class of such correlation lengths simply scale
with the spacing between homogeneously distributed, simultaneous plastic
events. This leads to a scaling of the correlation length with the shear 
rate as $\dot{\gamma}^{\nicefrac{-1}{2}}$ in 2D in the athermal
regime, regardless of the details of the model. The radius of the cooperative disk, defined as the near-field region
in which plastic events induce a stress redistribution that is not amenable to a mean-field treatment, notably follows this scaling. 
On the other hand,
the cooperative volume measured from the four-point stress susceptibility
and its dependence on the system size and the shear rate are
model-dependent.}
\vspace{0.5cm}
 \end{@twocolumnfalse}
  ]

\section{Introduction}

\footnotetext{\textit{$^{a}$~Univ. Grenoble Alpes, LIPhy, F-38000 Grenoble, France }}
\footnotetext{\textit{$^{b}$~CNRS, LIPhy, F-38000 Grenoble, France }}
\footnotetext{\textit{$^{c}$~ILM, Universit\'e de Lyon; UMR 5586 Universit\'e Lyon 1 et CNRS, F-69622 Villeurbanne, France}}
\footnotetext{\textit{$^{d}$~Institut Laue-Langevin, 6 rue Jules Horowitz, BP 156, F-38042 Grenoble, France }}

The onset of rigidity in a liquid cooled below its glass transition
temperature, as well as in granular matter packed more and more densely,
is accompanied by a growing, presumably diverging correlation length
\cite{Berthier2005,Heussinger2009}. To some extent, the situation
is similar to the onset of flow in an amorphous solid. Indeed, flow,
and the ensuing fluidisation, of the solid drives it away from the
critical elastic state that exists at vanishing shear rate \cite{Chen1991}. 
Diverging correlation lengths are then expected \cite{Lemaitre2009,Bocquet2009,Heussinger2010}
 when the shear rate goes to zero in the absence of thermal fluctuations \cite{Hentschel2010}. 

Extensive experimental research has been conducted to unveil the microscopic
details of the slow shear flow of these amorphous materials, from
the early works of Argon\cite{Argon1979} and Princen\cite{PRINCEN1985}
on bubble rafts and foams, to the more recent confocal microscopy
observations of colloids by Schall and co-workers\cite{Schall2007}
and the diffusive wave spectroscopy imaging of granular matter \cite{Amon2012}.
Numerical studies have also been largely contributed to our present understanding \cite{Maloney2006, Tanguy2006}.

It is now clear that the analogy with the glass (or jamming) transition
remains qualitative. In particular, while complex collective motion
on a large scale is observed around the jamming point, the onset of
flow in an amorphous solid is characterised by local rearrangements
of a handful of particles that induce a long-range elastic deformation
in the material, which may trigger new rearrangements in an avalanche-like
process (see Baret et al.~\cite{Baret2002} and references therein).

The enticing simplicity of this scenario has led to the emergence
of multiple models. A first class of models explicitly discard spatial
correlations and resort to a mean-field-like approach in which the
flow is described in terms of hops between {}``traps'' (metastable
configurations) that are facilitated by shear. The free volume theories
of Spaepen and others, the Shear Transformation Zone theory \cite{Falk1998,Langer2008},
the Soft Glassy Rheology model \cite{Sollich1997}, H\' ebraud and Lequeux's
equations \cite{Hebraud1998} all fall into this category, in spite
of the differences in the way they model the {}``traps'', or the
flow defects, and assess the hopping rates. In recognition of the
importance of flow heterogeneities, efforts have been made to extend
these theories beyond the homogenous, mean-field approximation. This
is generally achieved through the inclusion of a diffusive term in
the equations \cite{Fielding2009,Bocquet2009,Kamrin2012}. The diffusion-extended
equations have proved helpful in describing the striking manifestations
of spatial cooperativity in experiments \cite{Goyon2008a,Geraud2013,Kamrin2012}.

Still, one may nurture doubts about the adequacy of a simple diffusive
term in situations where heterogeneities interact via long-range (elastic)
interactions and fluctuations are large \cite{Jop2012}. This issue
is addressed by another line of modelling, namely lattice-based elastoplastic
models\cite{Baret2002,Picard2005,Vandembroucq2011}, pioneered by
Chen \& Bak \cite{Chen1991}, initially for the description of earthquakes,
and Argon \& Bulatov\cite{Bulatov1994,Bulatov1994a,Bulatov1994b}.
(Also see works by Homer \& Schuh\cite{Homer2009,Homer2010} for a similar, but off-lattice,
approach). However, it has been remarked\cite{Berthier2011} 
that the relevance of such models remains unclear owing to the vast
technical simplifications that they involve: generally, they are two-dimensional
(2D), they reduce the tensorial stress to a scalar quantity and neglect
the displacements of the elastoplastic blocks as the material is deformed.

In this contribution, we propose a detailed analysis of the importance
of the latter two aspects, namely the tensoriality of the stress and
convection, in an elastoplastic model. In particular, we shall quantitatively
probe the spatial correlations in the flow and the temporal fluctuations,
(both of which are omitted in purely mean-field approaches).

In Section 2, we clarify the general, continuum mechanics-based framework
of our model. We also show how convection can be implemented in 2D
mesoscopic models, and derive the relevant formulae for the propagators.
In order to best evidence the importance of the tensorial nature of
stress and the role of convection, we present a simple (but phenomenologically
rich) model in Section 3. The following section is dedicated to the
computation of diverse correlation lengths. Finally, in Section 5,
the probabilities involved in the model are refined so as to make
it more realistic in terms of the microscopic processes that have
been evidenced, and we assess how general the scalings we have derived
for the correlation lengths are.

\section{Description of the model}
\label{sec-model}

\subsection{\label{sub:General-framework}General framework}

The picture that emerged from the early experimental works on bubble
rafts of Argon\cite{Argon1979} and Princen\cite{Princen1983},\textbf{
}and that has since received ample confirmation from the observation
of diverse amorphous solids under slow shear\cite{Schall2007,Amon2012}
as well as numerical simulation of these systems\cite{Lemaitre2007,Tsamados2008},
revolves around localised rearrangements of particles bursting in
a mostly elastic medium, provided that the material is clearly solid
at rest, i.e., far enough from the glass transition or the jamming
point. On account of the shear geometry, these plastic events are
essentially tantamount to a relaxation of the local shear stress (via
particle rearrangement), although a transient, or even durable, local dilation\textbf{ }may occur simultaneously.

Let us first recall how this scenario can be interpreted in the framework
of continuum mechanics. More details can be found in  previous publications \cite{Nicolas2013a,Nicolas2013b}.

The elastic medium is characterised by incompressibility and linear
elasticity%
\footnote{Only very close to the onset of a plastic event is a significant departure
from linear elasticity observed\cite{Tsamados2010}.%
}, \emph{viz.}

\begin{equation}
\begin{cases}
\nabla\cdot u=0\\
\nabla\cdot\sigma^{\left(0\right)}-\nabla p^{\left(0\right)} & =0
\end{cases}\label{eq:initial_elasticity}
\end{equation}
where $u$ is the displacement field, $\sigma$ is the elastic stress tensor,
and $p$ is the pressure. If one introduces the linear strain tensor
$\boldsymbol{\epsilon}=\frac{\nabla u+^{t}\nabla u}{2}$, incompressibility
dictates that $\epsilon_{yy}=-\epsilon_{xx}$; this allows us to use
the following condensed notation, under the assumption of isotropy
of the medium: $\boldsymbol{\sigma}^{\left(0\right)}=2\mu\left(\begin{array}{c}
\epsilon_{xx}\\
\epsilon_{xy}
\end{array}\right)$, where $\mu$ is the shear modulus.

\selectlanguage{english}%
When a plastic event occurs in a region{\emph{
$S$}}, the system loses track of the reference elastic configuration
in this region, so that the material is locally fluidised. Region
$S$ is then dominated by dissipative forces $\sigma_\mathrm{diss}$, which
counter the relaxation of the stress $2\mu\epsilon_{\partial S}$
locally applied by the surrounding medium. For simplicity, we assume
that dissipation is linear in the shear rate, $\sigma_\mathrm{diss}=2\eta_\mathrm{eff}\dot{\epsilon}^{\mathrm{(pl)}}$,
with $\eta_\mathrm{eff}$ an effective viscosity, and we neglect the (subdominant)
elastic forces within $S$ for all the duration of the plastic event.
Then, force balance at the boundary $\partial S$ reads, in the absence
of inertia,
\begin{equation}
2\mu\epsilon_{\partial S}=2\eta_\mathrm{eff}\dot{\epsilon}^{\mathrm{(pl)}}.\label{eq:eps_deltaS}
\end{equation}
 The plastic strain $\dot{\epsilon}^{\mathrm{(pl)}}$ (per unit time) deforms
the boundary $\partial S$, thereby inducing an additional elastic
deformation $\dot{\epsilon}^{\mathrm{(1)}}$ in the medium. To leading order,
the increments of deformation $\dot{\epsilon}^{\mathrm{(1)}}(r)$ and pressure $\dot{p}^{\mathrm{(1)}}(r)$ 
in the matrix per unit time can be estimated by replacing the \emph{plastic} inclusion with an 
\emph{elastic} inclusion bearing an eigenstrain
\footnote{An eigenstrain $\epsilon^\star$ is defined by the following local elastic relation between stress and strain, $\sigma = 2\mu (\epsilon - \epsilon^\star)$.}
 (per unit time) equal to the plastic strain (per unit time) $\dot{\epsilon}^{\mathrm{(pl)}}$. It immediately follows
that,
\begin{equation}
2\mu\nabla\cdot\left(\dot{\boldsymbol{\epsilon}}^{\mathrm{(1)}}-\boldsymbol{\dot{\epsilon}}^{\mathrm{(pl)}}\right)-\nabla\dot{p}^{\mathrm{(1)}}=0\label{eq:Guillemette}
\end{equation}

Moving back to region $S$, since the plastic strain is a \emph{reaction}
to the elastic stress $2\mu\epsilon_{\partial S}$, it is expected
to lower the elastic strain at the boundary $\partial S$. On account
of the linearity of the problem, for a small inclusion $S$, one can
then write
\begin{equation}
\dot{\epsilon}_{\partial S}= \dot{\epsilon}_{\partial S}^\mathrm{(1)}
= -g_{0}\dot{\epsilon}^{\mathrm{(pl)}},\label{eq:eps_deltaS_dot}
\end{equation}
where $g_{0}$ is a positive scalar of order 1, whose precise value (in our implementation) will
be discussed in Section \ref{sec:model-presentation}.
The dynamics of the plastic event are obtained by combining
Eqs. \ref{eq:eps_deltaS} and \ref{eq:eps_deltaS_dot}, 
\begin{equation}
\dot{\epsilon}_{\partial S}=\frac{-g_{0}}{\tau}\epsilon_{\partial S},\label{eq:plastic_event_dynamics}
\end{equation}
where the timescale $\tau\equiv\nicefrac{\eta_\mathrm{eff}}{\mu}$ has been
introduced.

\subsection{Derivation of the elastic propagator}

\noindent
{\it 2.2.1 In an orthonormal frame}

We follow, and extend, the method proposed by Picard et al.~\cite{Picard2004}
to find the Green's function for Eq.~\ref{eq:Guillemette}, i.e., the
elastic propagator $\boldsymbol{\mathcal{G}}^{\infty}$.

With the shorthand $f$ for $-2\mu\nabla\cdot\boldsymbol{\dot{\epsilon}}^{\mathrm{(pl)}}$,
Eq.~\ref{eq:Guillemette} can be recast as
\[
2\mu\nabla^{2}\dot{u}^{\mathrm{(1)}}-\nabla\dot{p}^{\mathrm{(1)}} + f = 0,
\]
For convenience, we drop the (1)-superscripts denoting the
increments due to the plastic strain rate $\dot{\epsilon}^{\mathrm{(pl)}}$,
as well as the dots indicating time derivatives for the rest of this section:

\begin{equation}
2\mu\nabla^{2}u-\nabla p + f =0.\label{eq:Guillemette2}
\end{equation}
The combination of Eq.~\ref{eq:Guillemette2} with the incompressibility condition, $\nabla\cdot u=0$,
defines a well-known problem in hydrodynamics, for a (set of) pointwise source term(s) $f$. Its solution
 is most conveniently expressed in Fourier
coordinates $q\equiv (q_x, q_y)$ with the help of
the Oseen-Burgers tensor \cite{BarthesBiesel2010} $\mathcal{O}_{j}^{\, i}(q)=\frac{1}{\mu q^{2}}\left(\delta_{j}^{\, i}-\frac{1}{q^{2}}q_{j}q^{i}\right)$,
where $i$ and $j$ denote spatial directions, and we have written,
with Einstein's summation convention, %
\footnote{To simplify notations, we shall drop the hats for functions in Fourier space, 
that is, we shall write $u(q)$ instead of $\hat u(q)\equiv \iint dx\,dy\,u(x,y) e^{-\mathrm{i} (q_x x + q_y y)}$.}
$q^{2}\equiv q_{i}q^{i}$, $i \in \{x,y\}$, viz.,
\begin{equation}
u^{i}(q)=\mathcal{O}_{j}^{i}(q)\,f^{j}(q) = \frac{1}{2\mu q^{2}}\left(\delta_{j}^{i}-\frac{q^{i}q_{j}}{q^{2}}\right)f^{j}(q)\label{eq:displacement_non_orth}
\end{equation}
Finally, using {$\sigma_{j}^{\, i}(q)=2\mu\left[i\,\frac{q_{j}u^{i}+q^{i}u_{j}}{2}-\epsilon_{\, j}^{\mathrm{(pl)}\, i}(q)\right]$},
we arrive at
\[
\left(\begin{array}{c}
\sigma{}_{xx}\\
\sigma{}_{xy}
\end{array}\right) (q) = 2\mu\boldsymbol{\mathcal{G}}^{\infty}(q) \cdot \left(\begin{array}{c}
\epsilon^{\mathrm{(pl)}}_{xx}\\
\epsilon^{\mathrm{(pl)}}_{xy}
\end{array}\right) (q)
\]

where
\begin{equation}
\boldsymbol{\mathcal{G}}^{\infty}(q) \equiv\frac{1}{q^{4}}\left[\begin{array}{cc}
-(q_x^{2}-q_y^{2})^{2} & \,\,\,-2q_xq_y(q_x^{2}-q_y^{2})\\
-2q_x q_y(q_x^{2}-q_y^{2}) & -4q_x^{2}q_y^{2}
\end{array}\right].\label{eq:elastic_prop_orthonormal}
\end{equation}
Bear in mind that, under the assumption of incompressibility, $\epsilon_{xx}=-\epsilon_{yy}$.
In real space, the components of the elastic propagator $\boldsymbol{\mathcal{G}}^{\infty}$ display 
a four-fold angular symmetry and an $r^{-d}$ spatial decay, with $d$ the dimension of space, in accordance
with experimental and numerical evidence \cite{Schall2007,Puosi2014}. (See, e.g., Fig.1(right) in Ref.\cite{Martens2012} 
for a depiction of $\mathcal{G}^\infty_{22}$
in real space.)

It is worth noting that, in discretised space, with square mesh size
set to unity, only wavenumbers in the first Brillouin zone, \emph{viz.},
$q_x,q_y\in]-\pi,\pi]$, will be relevant. In addition, periodicity will
further restrict the nonzero Fourier modes to multiples of $\nicefrac{\pi}{L}$,
where $L$ is the periodic length in the direction under consideration.

\smallskip

\noindent
{\it 2.2.2 In a non-orthogonal frame}

As convection is to be included in the model, the initially orthonormal
frame $\left(x,y\right)$ will be deformed into a non-orthogonal frame
{$\left(x^{\prime},y^{\prime}\right)=\left(x-\gamma y,y\right)$},
where $\gamma$ is the average shear strain experienced by the cell,
so that the periodic replicas are advected with respect to each another
by the flow. In Fourier space, the correspondence between the wavenumbers
in the initial and deformed frames reads $(q_{x}^{\prime},q_{y}^{\prime})=(q_x,q_y+\gamma q_x)$. Note that
in the deformed frame covariant (e.g., $q_x^{\prime}$) and contravariant (e.g., $q^{\prime x}$) vector components need not be equal.
In Appendix \ref{sec:Derivation_of_prop_in_deformed_frame}, the derivation
of the elastic propagator $\mathcal{G}^{\infty}$ is extended to such
a non-orthogonal basis, with the help of the metric tensor in the
deformed frame. One arrives at an expression very similar to that
derived previously, Eq.~\ref{eq:elastic_prop_orthonormal}, where the
wavenumbers $q_x$ and $q_y$ in the orthonormal frame are simply replaced
by their expressions as functions of $q_x^{\prime}$and $q_y^{\prime}$,
\emph{viz.},
\[
\boldsymbol{\mathcal{G}}^{\infty} (q^\prime) \equiv\frac{1}{q^{4}}\left[\begin{array}{cc}
-(q_x^{\prime2}-q_y^{(\gamma)\,2})^{2} & \,\,\,-2q_x^{\prime}q_y^{(\gamma)}(q_x^{\prime2}-q_y^{(\gamma)\,2})\\
-2q_x^{\prime}q_y^{(\gamma)}(q_x^{\prime2}-q_y^{(\gamma)\,2}) & -4q_x^{\prime2}q_y^{(\gamma)\,2}
\end{array}\right].
\]

The shorthands $q_y^{(\gamma)}\equiv q_y^{\prime}-\gamma q_x^{\prime}$ and
$q^{4}\equiv\left(q_x^{\prime2}+q_y^{(\gamma)\,2}\right)^{2}$ have
been employed here.

This last formula brings to completion our effort to derive the propagator
for the stress redistribution in a uniform elastic matrix. We now
have to posit the rules for the local alternation of elastic regime
and plastic events, in light of the phenomenology evidenced experimentally
and numerically in the literature.

\section{Simplistic model\label{sec:Simplistic-model}}

\subsection{Presentation of the model\label{sec:model-presentation}}

Having dealt with the effect of a plastic event, we will now consider
the application of a finite strain rate to the material, with a velocity gradient along the
$y$-direction, so that the
time evolution of the local stress is a combination of the response
to the applied strain and the stress redistribution due to plastic
events, 
\begin{equation}
\partial_{t}\sigma\left(r,t\right)=\mu
\left(
\begin{array}{c}
0\\
\dot{\gamma}
\end{array}
\right)
+\int\mathcal{G}^{\infty}\left(r-r^{\prime}\right)\dot{\boldsymbol{\epsilon}}^{\mathrm{(pl)}}\left(r^{\prime},t\right)dr^{\prime},\label{eq:master_eq}
\end{equation}
where $\dot{\gamma}$ is the applied shear rate, and
 $\dot{\boldsymbol{\epsilon}}^{\mathrm{(pl)}}\left(r^{\prime},t\right)=\frac{\sigma\left(r^{\prime},t\right)}{2\mu\tau}$
if a plastic event is occurring locally (see Eq.~\ref{eq:plastic_event_dynamics}),
0 otherwise. Note that Eq.~\ref{eq:master_eq}
also applies to regions undergoing a plastic event, even though $\sigma$ is then of dissipative nature;
the local part of Eq.~\ref{eq:master_eq} then simply describes a Maxwell fluid of characteristic
time $\nicefrac{\tau}{g_{0}}$, where the value of the (positive) $g_0$ coefficient introduced in Eq.~\ref{eq:eps_deltaS_dot} 
is given by the (opposites of the) eigenvalues of the local component $\mathcal{G}^{\infty}(r-r^\prime= 0)$ in Eq.~\ref{eq:master_eq}. 
In our implementation of Eq.~\ref{eq:elastic_prop_orthonormal}, these eigenvalues are close to -0.5.

Turning to the criteria governing the onset and end of plastic events,
we first consider the very simple rules introduced by Picard \emph{et
al.} \cite{Picard2005} in a  scalar version of the model and their straightforward extension
 to the tensorial case (with $\epsilon_{xx}\neq 0$). Once the maximal shear stress
  $\left\Vert \sigma\right\Vert \equiv \sqrt{ \sigma_{xx}^2 + \sigma_{xy}^2}$
in a small region exceeds a given value $\sigma_{y}$, this region  has a finite
probability to yield. The associated yield rate is set to a constant,
$\tau_{liq}^{-1}$. We take $\tau_{liq}^{-1}=\tau^{-1}$, where $\tau$
is the characteristic time defined above. Besides, particle rearrangements
last for a constant (stress-independent) time $\tau_{res}$ on average. We choose units of time and stress such that $\tau=1$ and $\mu=1$.

Picard and co-workers showed that, in spite
of its simplicity, the model displays increasing complexity and cooperativity
as the shear rate $\dot{\gamma}$ tends to zero, while a mean-field-like
behaviour is recovered at large applied shear rates \cite{Picard2005}. In the following,
we measure diverse correlation lengths aimed at quantifying this cooperative
behaviour as $\dot{\gamma}$ decreases, and we assess to what extent
our results are altered by the insertion of a tensorial stress or/and
convection.

\subsection{Numerical implementation}

Before we proceed, a few words ought to be said about the numerical
implementation of the model. The system is discretised into a regular
square lattice of $N=L\times L$ elastoplastic blocks of unit size.
At each time step, the stress increments given by Eq.~\ref{eq:master_eq}
are computed in Fourier space, and then mapped back into real space.
For accuracy, we resolve the stresses on a finer mesh, in which
each elastoplastic block is made of four subcells. Also note that
the computation of the elastic propagator in discrete space may slightly
violate the equality of the streamline-averaged shear stresses imposed
by static mechanical equilibrium. To recover strict mechanical equilibrium,
we add a small \emph{ad hoc} offset to each streamline at every time step.
We checked that this procedure has only little impact on both the flow
curve and the correlation functions. After receiving
their stress increments, blocks may undergo a change of state, with
the probabilities given above. 

To account for convection, i.e., the advection of blocks along the
streamlines, the average shear deformation of the simulation cell
is updated at every timestep, \emph{viz.}, 
$\gamma_{total}= \dot\gamma t + (u_x^{top}-u_x^{bottom})/(y^{top}-y^{bottom})$,
where the non-affine displacements $u_x^{top}$ and $u_x^{bottom}$  of the
{}``top'' and {}``bottom'' streamlines,
with \emph{x} the flow direction, have been considered; note that these non-affine displacements
average to zero in an infinite system. Because the simulation cell is replicated periodically along both directions, 
one can always find
an appropriate $\gamma_{total}$ in the range $]\nicefrac{-1}{2},\nicefrac{1}{2}]$.
This defines the deformed frame. Since the flow is not strictly homogeneous,
or, equivalently, the deformation is not strictly affine, we must
additionally compute the displacement of each streamline so as to
be able to shift it adequately with respect to its neighbours. Details pertaining to the
calculation and implementation of this displacement are provided in Appendix \ref{sec:Displacement}.

\subsection{Flow curve and spatial organisation as a function of the restructuring
time\label{sub:Flow-curve-and-res-time}}

\begin{figure}[htpb]
\begin{centering}
\includegraphics[width=\columnwidth]{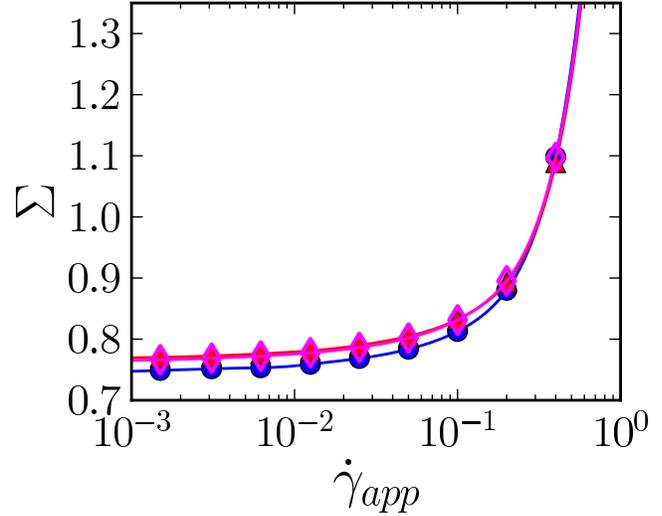}
\par\end{centering}

\caption{\label{fig:Flow-curves-Picard}Flow curves showing the macroscopic
shear stress $\Sigma$ as a function of the applied shear rate $\dot{\gamma}_{app}$,
for $\tau_{res}=1$. (\emph{Open diamonds}) static scalar model; (\emph{triangles})
static tensorial model; (\emph{dots}) convected tensorial model.}
\end{figure}

The study of the static (i.e., non-convected) scalar (i.e., $\sigma=\left(\begin{array}{c}
0\\
\sigma_{xy}
\end{array}\right)$) version of the model, as presented in Ref.\cite{Martens2012},
showed that at low enough shear rates, a transition from a (macroscopically)
homogeneous flow to permanent shear localisation occurs as the restructuring
time $\tau_{res}$ is increased, i.e., when it takes longer to the
material to {}``heal'' after a plastic event. Concomitantly with
the transition, a stress plateau develops in the flow curve. To what
extent is this scenario preserved when a tensorial stress is introduced
and convection  implemented?

First, we observe on Fig.\ref{fig:Flow-curves-Picard} that the flow
curve $\sigma\left(\dot{\gamma}\right)$ is hardly affected 
by the extension from a scalar to a tensorial stress; convection does
not alter it much either.

\begin{figure}[htpb]
\begin{centering}
\includegraphics[width=\columnwidth]{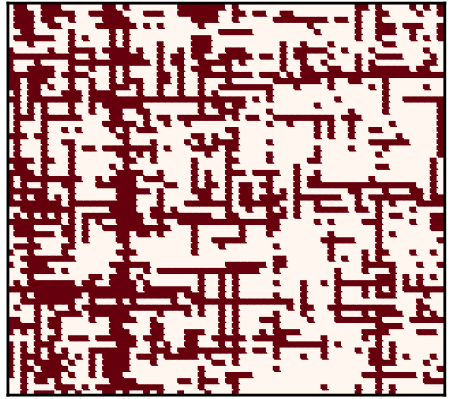}
\par\end{centering}

\caption{\label{fig:Binary_plastic_activity}Map of the plastic activity over
a strain window $\Delta\gamma=0.4$ in a system without macroscopic
shear localisation ($\tau_{res}=1$, $\dot{\gamma}=6.1\cdot10^{-3}$,
static tensorial model). Blocks that have yielded at least once over
the given strain window are shown in dark. The system is composed
of $64\times64$ blocks.}

\end{figure}

The extent of shear-localisation shall be quantified with the following
observable: $\kappa\left(\Delta\gamma\right)\equiv(n_{max}-n_{min})/(n_{max}+n_{min})$,
where $n_{max}$ and $n_{min}$ denote the maximum and minimum of
the line-averaged cumulated plastic activities over strain windows
$\Delta\gamma$ , i.e., the total time spent in the plastic state.
To smooth out fluctuations, line averages are further averaged with
the first neighbouring lines. With this definition, a vanishing value
of $\kappa$ signals homogeneous flow, whereas $\kappa=1$ indicates
full shear localisation. Note that even full shear localisation does
not preclude long-term diffusion of the bands, because they are not
pinned by a heterogeneity in the driving, as they would be in an experimental
Taylor-Couette geometry owing to the larger stress at the rotor. This
consideration highlights the necessity to keep $\Delta\gamma$ finite.
On the other hand, for small $\Delta\gamma$, spatial correlations
are always apparent, even in the absence of macroscopic shear localisation;
indeed, plastic events tend to align along {}``slip lines'', as
shown in Fig.\ref{fig:Binary_plastic_activity}, and consistently
with the molecular dynamics simulations reported in Ref \cite{Chaudhuri2013}.
Therefore, we choose a strain window of width $\Delta\gamma\approx10-30$,
after the (globally) stationary state has been reached. The qualitative picture
is robust to changes in $\Delta\gamma$.

\begin{figure}[htpb]
\begin{centering}

\subfloat[(\emph{Filled circles}) Scalar model.  (\emph{Open circles}) Static tensorial model.]
{  \includegraphics[width=0.5\columnwidth]{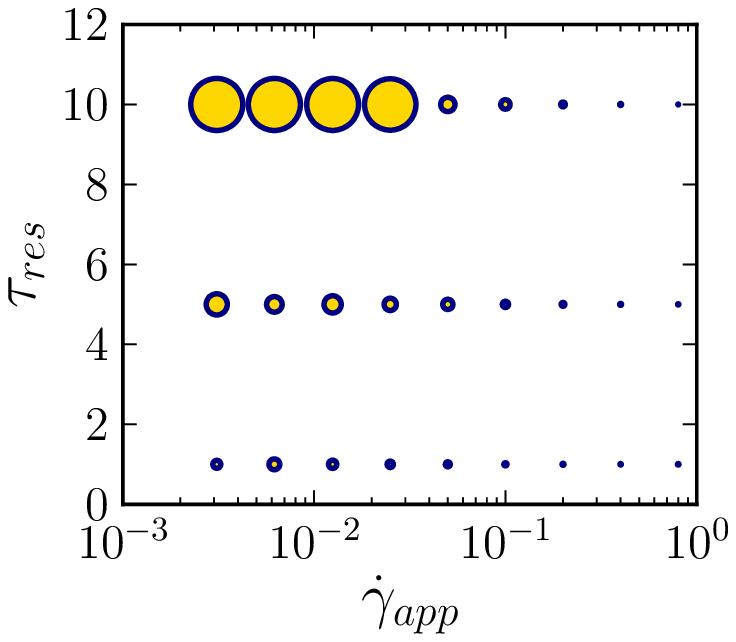}  }
\subfloat[(\emph{Filled circles}) Scalar model.  (\emph{Open circles}) Convected tensorial model.]
{\includegraphics[width=0.5\columnwidth]{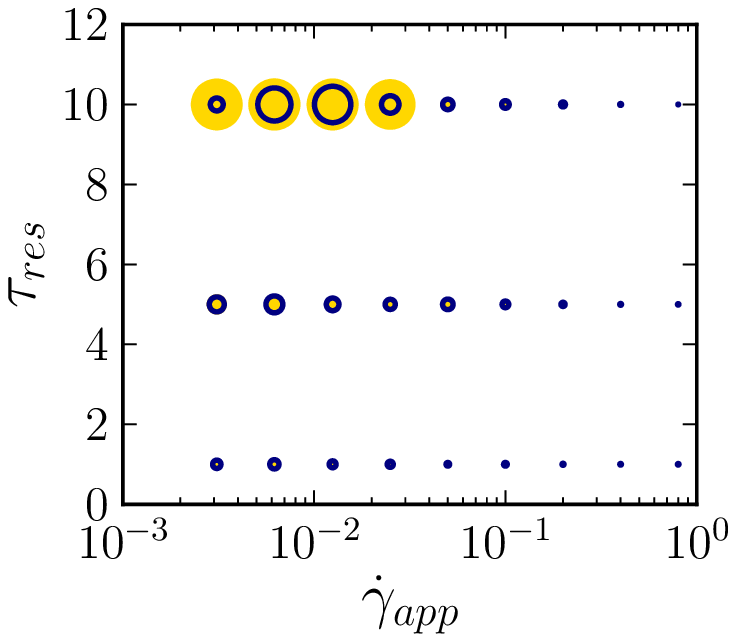} }

\par\end{centering}

\caption{\label{fig:Kappa observable}Dependence of $\kappa$ on the applied
shear rate $\dot{\gamma}_{app}$ and the restructuring time $\tau_{res}$. $\Delta\gamma=30$.
Circles are all the larger as the shear-banding observable $\kappa$
is (proportionally) large. The system consists of $128\times128$
blocks.}
\end{figure}

The values of the shear-banding observable $\kappa$ for various restructuring
times and applied shear rates are presented in Fig.\ref{fig:Kappa observable}.
Clearly, the flow is more prone to shear-banding at low applied shear
rates and long restructuring times. This is perfectly consistent with
our earlier findings in Ref.\cite{Martens2012} as well as with the
scenario described in Ref.\cite{Coussot2010}, whereby a long restructuring time
(plastic event) leads to a long-time decrease of the local stress, which
results in drastic shear-thinning on the macroscopic scale.
 The apparent decrease of $\kappa$ at very
low shear rates is most probably due to the diffusive motion of the
shear band, since the plastic activity is averaged over a fixed strain
window, i.e., increasingly large time windows as $\dot{\gamma}$ decreases.

A comparison between  the different versions of the model for a strain window $\Delta\gamma=30$
reveals that the inclusion of a tensorial stress in the static model
has virtually no effect on the shear-banding diagram (Figure \ref{fig:Kappa observable}a).
On the other hand, convection curtails shear-localisation to some
extent (Figure \ref{fig:Kappa observable}b), possibly because of the
enhancement of stress fluctuations outside the potential shear band,
which results in an increased mobility of the latter. The static \emph{vs.}
convected discrepancy vanishes when the strain window is reduced,
for instance, to $\Delta\gamma=5$ (\emph{data not shown}). For smaller
system sizes ($N=64\times64$), shear-banding profiles tend to be
more diffuse, and shear bands are more mobile, owing to larger fluctuations,
but the qualitative picture remains identical.

A major feature of the spatial organisation of the flow is left unnoticed
when considering only $\kappa$. Without convection, the streamwise
and crosswise directions are equivalent, because of the symmetry of
the stress tensor. Therefore shear bands are found equivalently in
either direction, which conflicts with experimental observations.
As expected, enforcing convection breaks the symmetry and only allows
shear bands in the flow direction.

\begin{figure}[htpb]
\begin{centering}
\subfloat[Without convection]{\includegraphics[width=0.5\columnwidth]{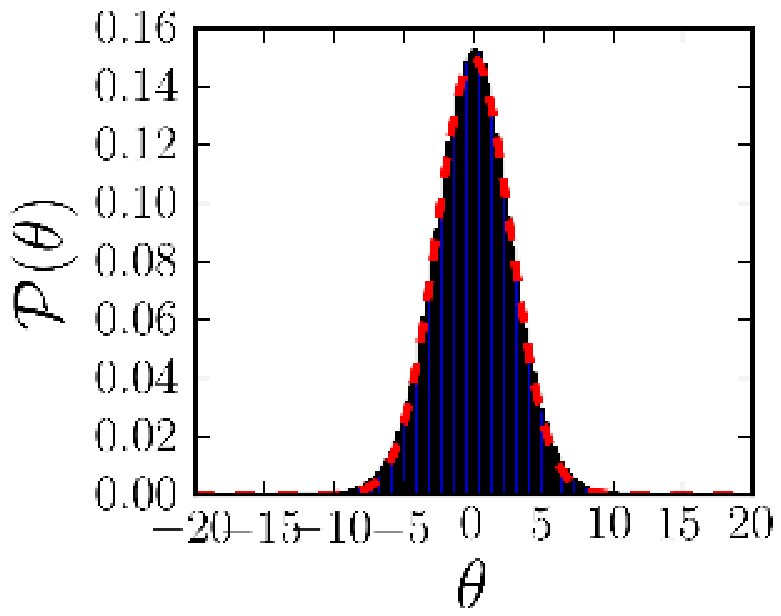}
}\subfloat[With convection]{\includegraphics[width=0.5\columnwidth]{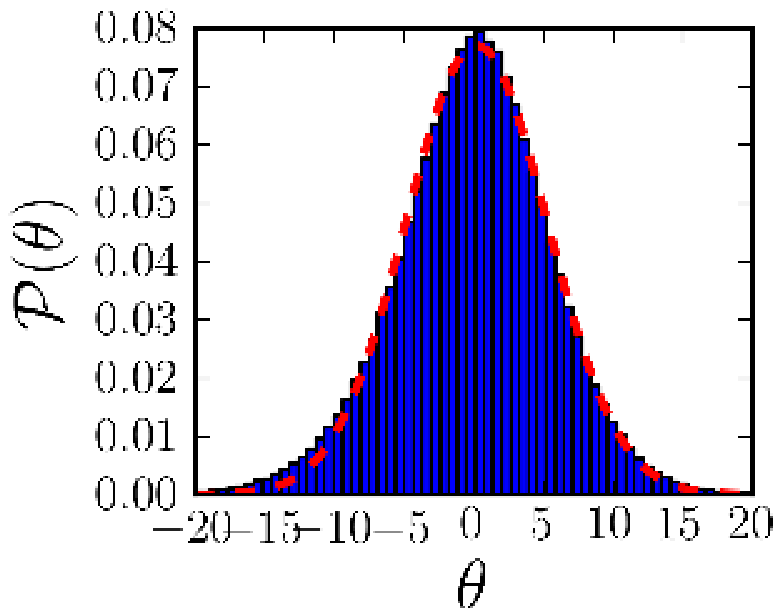}
}
\par\end{centering}
\label{Ptheta}
\caption{\label{fig:Ptheta}Distribution of yielding angles $\theta$ in degrees for
$\tau_{res}=1$ for tensorial models with and without convection}

\end{figure}

The growth of cooperativity with increasing restructuring times $\tau_{res}$
is also reflected by the distribution of principal directions of plastic
events. Let $\theta\in\left[-90^{\circ},90^{\circ}\right]$ be the
corresponding angle with respect to the macroscopic shear (xy) direction,
that is, $\cos\left(2\theta\right)=\frac{\sigma_{xy}}{\sigma},\,\sin\left(2\theta\right)=\frac{-\sigma_{xx}}{\sigma}$,
where $\sigma\equiv\sqrt{\sigma_{xx}+\sigma_{xy}}$. In the absence
of cooperativity, one expects plastic events to be aligned with the
applied shear, hence $\theta=0$. Cooperativity broadens the distribution
$\mathcal{P}\left(\theta\right)$. Indeed, as $\tau_{res}$ increases
from 1 to 10 time units, the standard deviation of the distribution
approximately doubles, at a given shear rate. It is also worth noting
that switching on convection also results in the doubling of the standard
deviation of the distribution, as shown in Fig. \ref{fig:Ptheta}. Once again, we ascribe this to the
enhancement of fluctuations due to convection.

In the following section, we come back to the original case $\tau_{res}=1$.

\section{Correlation lengths\label{sec:Correlation-lengths}}

\subsection{Four-point susceptibility $\chi_{4}$}

In glassy systems, instantaneous one-point observables hardly differ
from their counterparts in the fluid state, and the search for an
observable whose static correlations would distinguish the two states
has not borne much fruit so far. On the other hand, time correlations
of local observables have proven of great use as order parameters
\cite{Toninelli2005}. Here, we study the stress autocorrelation
function $c\left(r,\Delta t\right)\equiv\delta\sigma\left(r,0\right)\delta\sigma\left(r,\Delta t\right)$
, where $\delta\sigma\equiv\sigma-\bar{\sigma}$. Spatial correlations
are probed with the four-point correlator
\begin{equation}
\mathcal{G}_{4}\left(\Delta r,\Delta t\right)\equiv\left\langle c\left(O,\Delta t\right)c\left(\Delta r,\Delta t\right)\right\rangle -\left\langle c\left(O,\Delta t\right)\right\rangle ^{2},
\end{equation}
where the brackets denote an average over time, or, equivalently,
configurations (since the system is stationary). Note that the above
definition is independent of the choice of origin $O$.

The precise definition of $c\left(r,\Delta t\right)$ deserves a comment
in presence of convection, in which case blocks may move over $\Delta t$.
In line with the definition of $c$ as the stress autocorrelator,
we adopt a Lagrangian description and compute $c$ as $\left\langle \delta\sigma\left(r,0\right)\delta\sigma\left(r^{\prime},\Delta t\right)\right\rangle $,
where $r^{\prime}$ is the convected position at $\Delta t$ of the
block that was initally at position $r$. Note that the same idea
prevailed in Furukawa et al.'s definition \cite{Furukawa2009} of
the four-point susceptibility of a system under shear.

\begin{figure}[htpb]
\begin{centering}
\subfloat[Static tensorial model. ]{\includegraphics[width=0.5\columnwidth]{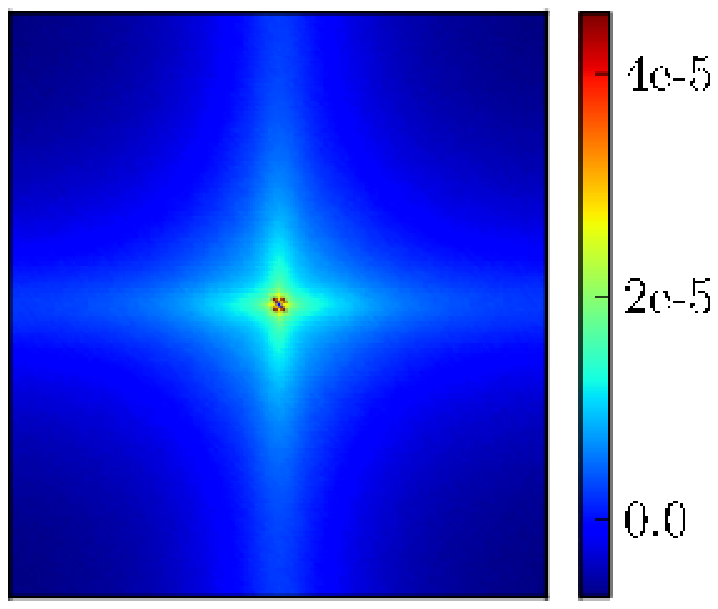}

}\subfloat[Convected tensorial model. $\mathcal{G}_{4}$ is represented as a
function of the \emph{initial} positions of the convected blocks. ]{\includegraphics[width=0.5\columnwidth]{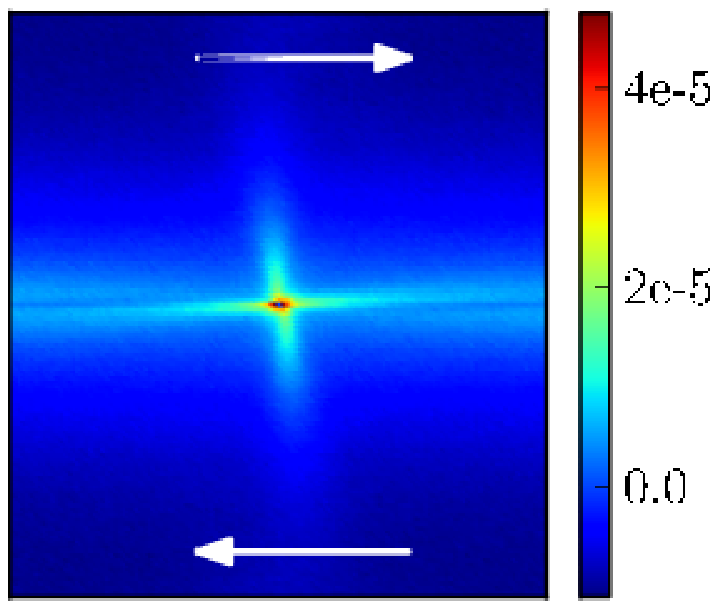}

}
\par\end{centering}

\caption{\label{fig:G4_profile}Spatial profile of the four-point correlator
$\mathcal{G}_{4}$ at $\Delta\gamma=0.37\approx\Delta\gamma^{\star}$,
$\dot{\gamma}_{app}=10^{-3}$. System size: $128\times128$. Because
of the comparatively very large value of the stress autocorrelator
$\mathcal{G}_{4}(0,\Delta t$), the central cell has been artificially
coloured.}
\end{figure}

Figure \ref{fig:G4_profile} shows the spatial profile of $\mathcal{G}_{4}$
at $\dot{\gamma}_{app}=10^{-3}$ for a delay time $\Delta t=0.37$
of the order of the stress autocorrelation time. The profiles for
the static versions of the model are indistinguishable with the naked
eye, and remain identical if one substitutes $\sqrt{\sigma_{xx}+\sigma_{xy}}$
for $\sigma_{xy}$ in the definition of the time correlator $c$.
They display long branches in the velocity and velocity gradient directions,
in accordance with the directions of the positive lobes of the $xy$-component of the elastic
propagator $\mathcal{G}^{\infty}$. The large spatial extent of these
branches is in part due to the periodicity of the system in the two
directions. 

Adding convection radically changes the picture. Most notably, the
symmetry between the $\left(Ox\right)$ (i.e., flow) and $\left(Oy\right)$
(i.e., velocity gradient) directions is broken. The streamline going
through the origin keeps a forward-backward ($x\rightarrow-x$) symmetry,
but outside this line no such symmetry is preserved. In particular,
the branche approximately along $\left(Oy\right)$ direction is tilted,
so that a block initially located at position $-x$ in this branch
will be convected to position $x$ after the lag strain $\Delta\gamma$,
meanwhile passing through the $\left(Oy\right)$-lobe of the stress
propagator. The distinction between the generic features of $\mathcal{G}_{4}$
and those specific to the present model shall be addressed in Section
\ref{sub:4-point-susceptibility-refined}.

The integral of $\mathcal{G}_{4}$ over space, at fixed $\Delta t$,
yields the four-point susceptibility $\chi_{4}$, that is, the variance
of the two-time correlation function with time, $\chi_{4}\left(\Delta t\right)=V\,\cdot\mathrm{Var}\left(C\left(\Delta t\right)\right)$,
where $V$ is the volume of the system, $C(\Delta t)\equiv V^{-1}\int c\left(r,\Delta t\right)dr$,
and the variance operator $\mathrm{Var}$ has its usual definition,
$\mathrm{Var}\left(\cdot\right)\equiv\left\langle \cdot^{2}\right\rangle -\left\langle \cdot\right\rangle ^{2}$.
If the integral is normalised by the value at the origin\cite{Toninelli2005},
viz., $\tilde{\chi}_{4}(\Delta t)\equiv\nicefrac{\chi_{4}(\Delta t)}{\mathcal{G}_{4}\left(0,\Delta t\right)}$,
it then gives an estimate of the spatial volume in which the stress
evolves in a correlated fashion with that at the origin. To illustrate
this schematically, suppose that the system consists of $\nicefrac{V}{V_{coop}}$
entirely correlated, but mutually decorrelated, regions of volume
$V_{coop}$ each. A simple application of the central limit theorem
yields 
\begin{eqnarray}
\tilde{\chi}_{4}(\Delta t) & = & \frac{V\,\cdot\mathrm{Var}\left(C\left(\Delta t\right)\right)}{\mathcal{G}_{4}\left(0,\Delta t\right)}\\ \nonumber
 & \approx & V_{coop}\frac{\mathrm{Var}\left(c\left(O,\Delta t\right)\right)}{\mathcal{G}_{4}\left(O,\Delta t\right)}\\ \nonumber
 & \approx & V_{coop}.
\end{eqnarray}
It follows that the peak $\tilde{\chi}_{4}^{\star}$ of $\tilde{\chi}_{4}(\Delta t)$,
which is reached at a lag time $\Delta t^{\star}$ close to the stress
autocorrelation time, is a measure of the maximal cooperativity in
the flow. Here, $\Delta t^{\star}$ is such that $\dot{\gamma}\Delta t^{\star}\approx0.3-0.5$
is of the order of the yield strain. The value of $\mathcal{G}_{4}\left(O,\Delta t^{\star}\right)$
depends even less on the shear rate.

Now, we turn to a more detailed analysis of the variations of the
cooperative volume $\tilde{\chi}_{4}^{\star}$ with the applied shear
rate $\dot{\gamma}$, starting with the static models. At rather
high shear rates, $\tilde{\chi}_{4}^{\star}$ is independent of the
system size and exhibits the following shear rate dependence:
\begin{equation}
\tilde{\chi}_{4}^{\star}\sim\dot{\gamma}^{-\beta},
\end{equation}
with $\beta\approx0.9$ for both the scalar and the tensorial
models. When the shear rate is decreased, the cooperative volume increases,
and finally saturates at a value proportional to $L^{\nicefrac{3}{2}}$
when the whole simulation cell becomes correlated. The transition
takes place around a shear rate $\dot{\gamma}_{c}$ such that $\dot{\gamma}_{c}^{-\beta}\sim L^{\nicefrac{3}{2}}$.
Therefore, following Ref.  \cite{Martens2011}, we propose the
scaling
\begin{equation}
\tilde{\chi}_{4}^{\star}\sim L^{\nicefrac{3}{2}}f\left(\dot{\gamma}^{-\beta}L^{-\nicefrac{3}{2}}\right),
\end{equation}
where $f\left(x\right)\sim x$ when $x\rightarrow0$ and $f\left(x\right)\sim1$
when $x\rightarrow\infty$. Figure \ref{fig:chi4_vs_gdot} shows that
a nice collapse can then be achieved.

Using the fractal dimension $\nicefrac{3}{2}$ for the cooperative
region, one can assess the four-point correlation length, $\xi_{4}\sim\tilde{\chi}_{4}^{\star\nicefrac{2}{3}}\sim\dot{\gamma}^{\nicefrac{2\beta}{3}}$.
Interestingly, the exponent $\nicefrac{2\beta}{3}\approx0.6$, for
both scalar and tensorial models, is close to the exponent
$\nicefrac{1}{2}$ extracted by Lema\^itre and Caroli\cite{Lemaitre2007}
from the transverse diffusion coefficient in their 2D molecular dynamics
simulations (although, admittedly, they found linear avalanches in
2D, instead of our $\nicefrac{3}{2}$ fractal exponent). On the other
hand, it differs from the exponent $\nicefrac{\ensuremath{1}}{4}$
predicted by the kinetic elastoplastic theory of Bocquet et al. \cite{Bocquet2009}.
More surprisingly, it also differs from the exponent reported in Ref.  \cite{Martens2011}
for a slightly different rescaling of the observable, but with a model
identical to the present one. We have checked that the scaling proposed
in  Ref. \cite{Martens2011} provides a poorer fit to our more extensive data set
(see  Fig. \ref{fig:scalingtest}).

The insertion of convection modifies the scaling thoroughly. Consistently
with the  atomistic simulations of 
Maloney and Lema\^itre\cite{Maloney2004}, and Lema\^itre and Caroli\cite{Lemaitre2009},
linear correlations (referred to as {}``slip lines'' by Maloney
and Lema\^itre, see Fig.\ref{fig:G4_profile}) then dominate and $\tilde{\chi}_{4}^{\star}$
saturates at a value apparently almost linear in $L$ (see Fig.\ref{fig:chi4_CBP}).
The non-saturated regime in which the cooperative volume depends solely
on the shear rate, is never truly reached in our simulations~:
finite-size effects are always dominant, which hampers our search
for a scaling law.

\begin{figure}[htpb]
\begin{centering}
\includegraphics[width=\columnwidth]{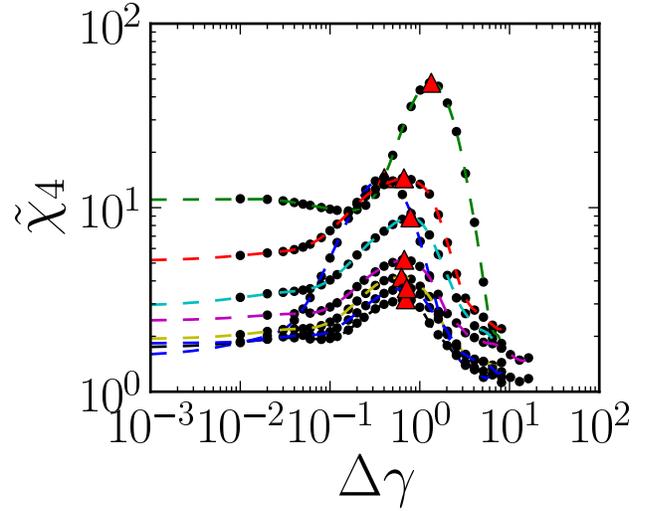}
\par\end{centering}

\caption{Four-point susceptibility as a function of strain delay
$\Delta\gamma=\dot{\gamma}\Delta t$, for various shear rates (increasing
$\dot{\gamma}$ from top to bottom). The red triangles indicate the
maximal values.}
\end{figure}

\begin{figure}[htpb]
\subfloat[Static scalar model]{
\includegraphics[width=0.5\columnwidth]{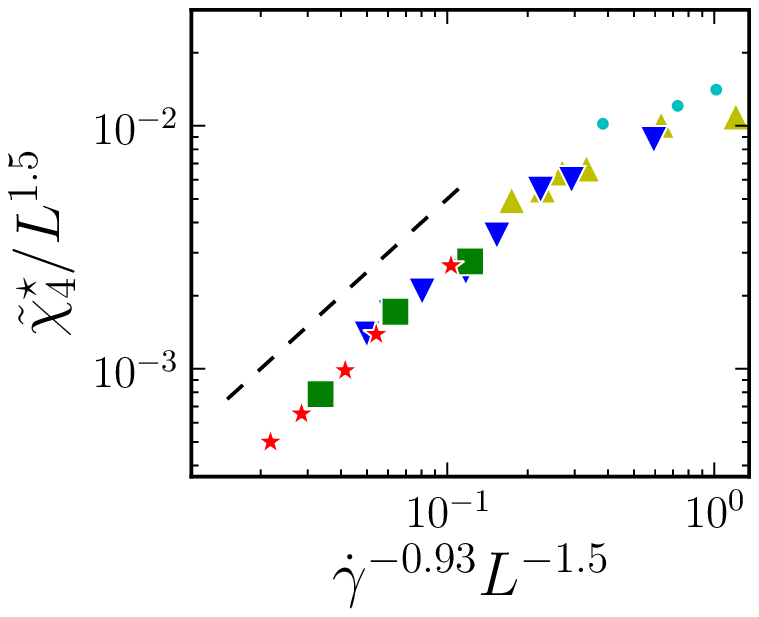}
}
\subfloat[Static tensorial model]{
\includegraphics[width=0.5\columnwidth]{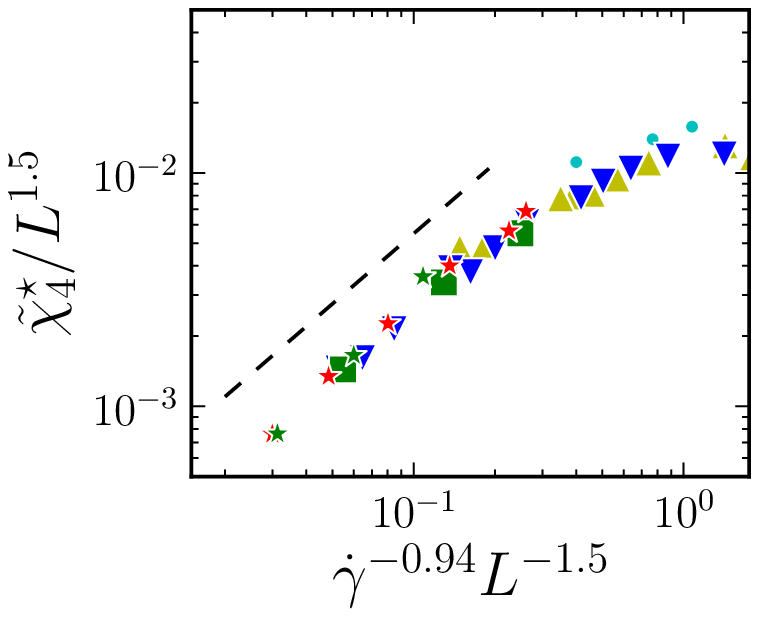}
}


\caption{\label{fig:chi4_vs_gdot}Scaling of the maximal cooperative volume
$\tilde{\chi}_{4}^{\star}$ as a function of shear rate $\dot{\gamma}$ in the static models. The vertical axis
is $L^{-1.5}\chi_{4}^{\star}$, and the horizontal axis is the rescaled
shear rate $\dot{\gamma}^{-\beta}L^{-1.5}$, with $\beta=0.93$ for
the scalar model and $\beta=0.94$ for the tensorial model. Various
linear sizes of the (square) system are studied: $L=$ (\emph{cyan
dots}) $32$, (\emph{yellow} \emph{dots}) $64$, (\emph{blue} \emph{triangles})
$128$., (\emph{green} \emph{squares}) $192$, (\emph{red} \emph{stars})
$256$, (\emph{green} \emph{stars})
$384$. As a guide to the eye, we have plotted a dashed line with
slope 1.}
\end{figure}

\begin{figure}[htpb]
\begin{centering}
\includegraphics[width=\columnwidth]{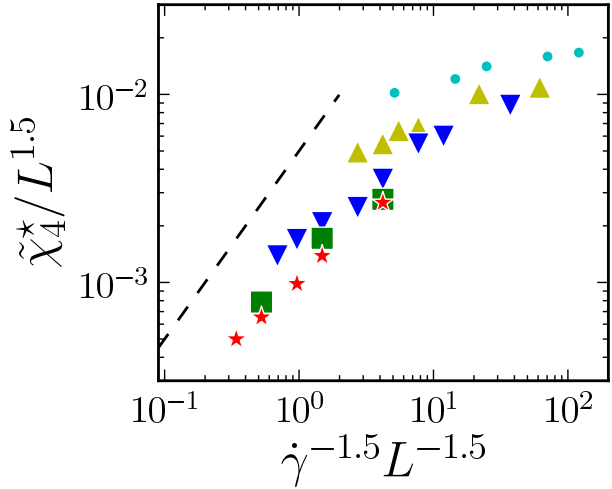}
\par\end{centering}
\caption{\label{fig:scalingtest}
Test of the scaling of $\tilde{\chi}_{4}$
proposed in \cite{Martens2011}, viz., $\frac{\tilde{\chi}_{4}}{L^{\nicefrac{3}{2}}}=f\left(\frac{\dot{\gamma}^{-\nicefrac{3}{2}}}{L^{\nicefrac{3}{2}}}\right)$,
for the static scalar model. Various linear sizes of the (square)
system are studied: $L=$ (\emph{cyan dots}) $32$, (\emph{yellow}
\emph{dots}) $64$, (\emph{blue} \emph{triangles}) $128$., (\emph{green}
\emph{squares}) $192$, (\emph{red} \emph{stars}) $256$. As a guide
to the eye, we have plotted a dashed line with slope 1. Data have
been averaged over $\Delta\gamma\approx300.$}
\end{figure}

\begin{figure}[htpb]
\begin{centering}
\includegraphics[width=\columnwidth]{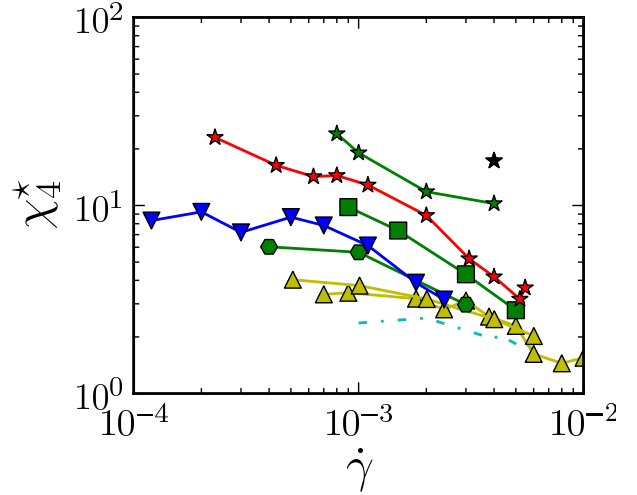}
\par\end{centering}

\caption{\label{fig:chi4_CBP}Maximal cooperative volume $\tilde{\chi}_{4}^{\star}$
in the convected tensorial model.
Various linear sizes of the (square) system are studied: $L=$ (\emph{cyan
dash-dotted line}) 32, (\emph{yellow triangles}) 64, (\emph{green hexagons})
96, (\emph{blue triangles}) 128, (\emph{green squares})
192, (\emph{red stars}) 256, (\emph{green stars}) 384,
(\emph{black star}) 512.}
\end{figure}

\subsection{Cooperative disk}

In this section, we propose an alternative protocol to define a correlation
length of the system, rooted in the interpretation of the onset of
flow in an amorphous solid as a dynamic phase transition\cite{Bocquet2009}.
Setting the macroscopic shear stress $\sigma_{xy}$ as a control parameter,
we view the steady-state strain-rate tensor $\dot{\boldsymbol{\epsilon}}$
as an order parameter, which goes to zero below the yield stress and
continuously increases above it.

One may then wonder whether a mean-field approach is applicable, or
whether it breaks down because of (spatiotemporal) fluctuations. To answer this question, we assess how large the standard
deviation of the fluctuations $\sqrt{\left\langle \left\Vert \boldsymbol{\delta\dot{\epsilon}}\right\Vert {}^{2}\right\rangle }$
experienced at one point $M$ in the system is, compared to the mean
value $\left\Vert \left\langle \boldsymbol{\dot{\epsilon}}\right\rangle \right\Vert $
of the order parameter. Except at very large shear rates, this ratio
is always large, because plastic events occurring close to $M$ cause
very large fluctuations. But should we only consider the effect of
\emph{distant} plastic events, would the fluctuations then be negligible,
and a mean-field treatment applicable for them? Concretely, at arbitrary
points, we compute the mechanical noise $\boldsymbol{\dot{\epsilon}}\left(\xi\right)$
due to plastic events taking place farther than some distance $\xi$
from $M$. The use of a Ginzburg-Landau criterion $\nicefrac{\sqrt{\left\langle \left\Vert \boldsymbol{\delta\dot{\epsilon}}\right\Vert {}^{2}\right\rangle }}{\left\Vert \left\langle \boldsymbol{\dot{\epsilon}}\right\rangle \right\Vert }\left(\xi\right)<1$
allows us to distinguish, for any point $M$ in the system, a cooperative
disk of radius $\xi^{\star}$, from an outer region which is amenable
to a mean-field treatment, i.e., which satisfies the criterion. With regard to the \emph{instantaneous} mechanical
noise at $M$, the details of the individual plastic events occurring
within the cooperative disk will matter, whereas outside the disk
they will not. 

In addition, the comparison between the cooperative length $\xi^{\star}$
and the size of a structural rearrangement (the unit size, here) will
be a valuable hint as to whether our model gives credence to mean-field
analyses \cite{Dahmen2009}, possibly complemented with a diffusion
term to account for spatial fluctuations \cite{Bocquet2009,Fielding2009,Kamrin2012}.

\begin{figure}[htpb]
\begin{centering}
\includegraphics[width=\columnwidth]{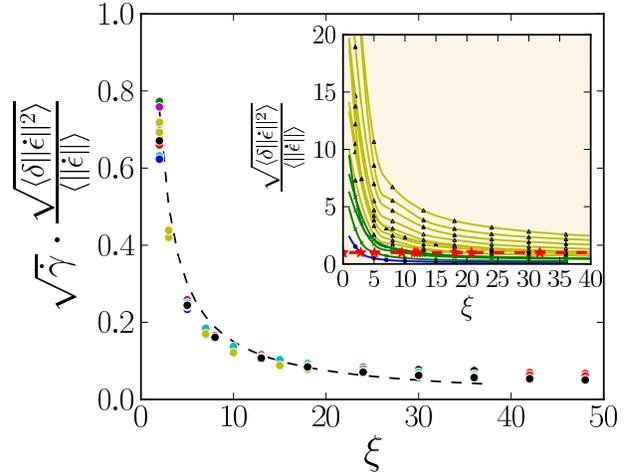}
\par\end{centering}

\caption{\label{fig:Xi_Curves_CBP}Ratio of fluctuations over mean value of
the mechanical noise due to plastic events taking place farther than
$\xi$, rescaled with $\sqrt{\dot{\gamma}}$, as a function of $\xi$,
in the convected tensorial model. System size: $L=256$. Data include
points at $\dot{\gamma}=5.4\cdot10^{-4},\,8.1\cdot10^{-4},\,1.2\cdot10^{-3},\,1.8\cdot10^{-3},\,2.7\cdot10^{-3},\,3.5\cdot10^{-3}$,
$\,4.0\cdot10^{-3},\,6.0\cdot10^{-3},\,9.0\cdot10^{-3},\,1.3\cdot10^{-2},\,1.4\cdot10^{-2},\,2.0\cdot10^{-2},\,1.3\cdot10^{-2},$$2.0\cdot10^{-2},\,5.6\cdot10^{-2},2.0\cdot10^{-2},\,0.23$.
The dashed black line represents $y=\nicefrac{1.5}{\xi}$.\emph{ Inset:
}Same data, not rescaled with $\sqrt{\dot{\gamma}}$. The yellow ($\dot{\gamma}<10^{-2}$),
green ($10^{-1}<\dot{\gamma}<10^{-2}$), and blue ($\dot{\gamma}>10^{-1}$)
lines are guides to the eye. The dashed red line marks the limit $\frac{\sqrt{\left\langle \left\Vert \boldsymbol{\delta\epsilon}\right\Vert {}^{2}\right\rangle }}{\left\Vert \left\langle \boldsymbol{\epsilon}\right\rangle \right\Vert }=1$. }

\end{figure}

Figure \ref{fig:Xi_Curves_CBP} shows that the data collapse onto
a master curve,

\begin{equation}
\frac{\sqrt{\left\langle \left\Vert \boldsymbol{\delta\dot{\epsilon}}\right\Vert {}^{2}\right\rangle }}{\left\Vert \left\langle \boldsymbol{\dot{\epsilon}}\right\rangle \right\Vert }\left(\xi\right)\sim\frac{1}{\xi\sqrt{\dot{\gamma}}}.\label{eq:xi_star_scaling}
\end{equation}

\begin{figure}[htpb]
\begin{centering}
\includegraphics[width=\columnwidth]{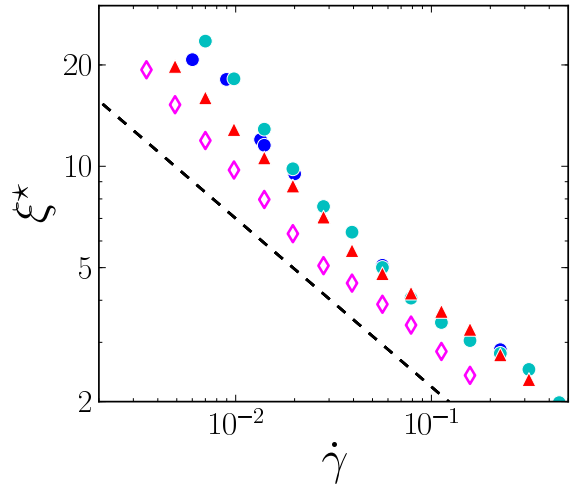}
\par\end{centering}

\caption{\label{fig:Xi_star_vs_gammadot}Dependence of the cooperative length
$\xi^{\star}$ on the applied shear rate $\dot{\gamma}_{app}$. The
dashed line has slope $\nicefrac{-1}{2}$. (\emph{Open diamonds})
static scalar model; (\emph{triangles}) static tensorial model; (\emph{dots})
convected tensorial model. Linear system size: $L=128$, except for
dark blue dots ($L=256$). }

\end{figure}

We have checked that this scaling is not marred by finite-size effects.
It immediately follows from Eq.~\ref{eq:xi_star_scaling} that $\xi^{\star}\sim\dot{\gamma}^{\nicefrac{-1}{2}}$,
which is confirmed by Fig. \ref{fig:Xi_star_vs_gammadot} for all
versions of the model. The assumption that plastic events should be
only weakly interacting in a slow flow, at low temperature (as expressed
in ref \cite{Langer2008} and more generally in mean-field-like approaches)
may therefore seriously be called into question. An  analysis
of the impact of these instantaneous fluctuations on the yielding
rates is presented in Ref.\cite{Nicolas2014}.

Although the large values of $\xi^{\star}$ point to the sensitivity
to plastic event details over a large region, a simple calculation
discarding static spatial correlations between plastic events already
provides a satisfactory explanation of the scaling behaviour of $\xi^{\star}$,
Eq.~\ref{eq:xi_star_scaling}. Indeed, under the assumption of randomly
located plastic events, we recover the desired scaling law, Eq. \ref{eq:xi_star_scaling},
as detailed in Appendix \ref{sec:Estimation-of-xi_star}. The derivation
is based on the following: the typical mechanical noise $\left\Vert \boldsymbol{\dot{\epsilon}}\right\Vert $
created by a plastic event at a distance $r$ amounts to $\frac{\dot{\epsilon}^{pl}}{r^{d}}$
in \emph{d }dimensions, whereas its mean value, for all possible relative
positions, is only of order $\frac{\dot{\epsilon}^{\mathrm{(pl)}}}{L^{d}}$,
because of the compensation between the positive and negative lobes
of the elastic propagator; lastly, the average number of simultaneous
plastic events is proportional to the shear rate. We would like to
emphasise that the scope of the derivation extends far beyond the
present model;\emph{ in fine}, we simply find that the cooperative
length scales with the spacing between homogeneously-distributed,
simultaneous plastic events. In other words, $\xi^{\star}$ is not
sensitive to the presence of correlated {}``slip lines'' in the
flow and, more generally, other deviations from a homogeneous distribution
of plastic events.

It is therefore not a surprise to find the same scaling of the correlation
length as that used by Ref.\cite{Lemaitre2009} to interpret the transverse
diffusivity in their molecular dynamics simulations, namely a dependence
on $\dot{\gamma}^{\nicefrac{-1}{d}}$. Note that the authors of Ref.\cite{Lemaitre2009}
had rationalised it by secluding the non-overlapping near-field {}``flips''
(plastic events) from an incoherent background of {}``flips''. In
many respects, our cooperative disk approach comes in the wake of
theirs.

\subsection{Picard's crossover method}

We would like to add a word about the characteristic length introduced
by Picard \emph{et al.} in Ref. \cite{Picard2005}, namely the linear
size of the system below which the macroscopic stress-drops, normalised
by the average stress, saturate due to finite-size effects. It seems
to us that this saturation occurs when the applied shear rate is decreased
to such an extent that there is at most one plastic event in the simulation
cell at any time, in which case the characteristic length also provides
a measure of the distance between plastic events, in a somewhat cruder
way than the cooperative length $\xi^{\star}$ introduced previously.
Accordingly, Picard's correlation length, assessed in the static scalar
model, was also reported to scale with $\dot{\gamma}^{-\nicefrac{1}{2}}$.

As a partial summary of the results collected with this first model,
we have considered different correlation lengths; all were found to
be approximately proportional to the inverse square root of the shear
rate, with the exception of the four-point correlation length in the
convected system, whose scaling remains elusive for us. How universal
is the $\dot{\gamma}^{\nicefrac{-1}{2}}$- scaling (in 2D)? Spurred
on by this question, we propose a refined model, featuring somewhat
more realistic dynamical rules.

\section{Refinement of the model}

The model studied in the previous section does not allow us to recover
the Herschel-Bulkley flow curve, $\sigma=\sigma_{0}+A\dot{\gamma}^{n}$,
with $n\approx0.5$, very often reported in experiments as well as
in atomistic simulations,
even for athermal materials. Here, we devise a model dedicated mostly
to the latter type of materials, insofar as thermal activation of plastic
events will be precluded; nevertheless, an extension to thermal fluctuations
and ageing in glasses shall also be touched upon.

\subsection{Onset of a plastic event}

The delay $\tau_{liq}$ before an elastoplastic block yields after
crossing the yield stress, as introduced above, is questionable from
a potential energy landscape (PEL) perspective. (We must however mention
that, to some extent, it has been rationalised in the context of spring-and-dashpot
models on a periodic pinning potential, see for instance \cite{Marmottant2013}).
Consequently, we replace this criterion with a traditional yield criterion%
\footnote{Note that the distinction between Tresca and von Mises yield criteria
breaks down in 2D, both criteria being equivalent for $d\leqslant2$.%
}, whereby a block yields as soon as the yield stress is exceeded.

Instead of a single yield stress, a distribution of yield stresses
is introduced, on the following basis: let us consider the potential
energy landscape and coarse-grain it (for practical reasons) so that
too shallow energy basins, of depth $E_{y}<E_{min}$, are discarded.
For $E_{y}\geqslant E_{min}$, following the Soft Glassy Rheology
model, we choose an exponential distribution of energy barriers, so
that,
\begin{equation}
\rho\left(E_{y}\right)\propto\begin{cases}
\exp\left(\frac{E_{min}-E_{y}}{\left\langle E_{y}\right\rangle }\right) & \text{ if }E_{y}\geqslant E_{min}\\
0 & \text{ otherwise},
\end{cases}
\end{equation}
where the mean value $\left\langle E_{y}\right\rangle $ is adjusted
so that mean yield strain takes the realistic value $\left\langle \gamma_{y}\right\rangle =0.1$,
for a two-dimensional material.
At the end of every plastic event, a new energy $E_{y}$ is randomly
assigned to the block from the $\rho$ distribution. 
Regarding the duration of a plastic event, we make the coarse approximation
that, while being sheared, there is a typical distance (measured in
terms of local strain) between two successively visited metabasins.
Of course, this distance is related to the fineness of the coarse
graining, i.e., $E_{min}$. Consequently, we suggest that a plastic
event ends when a total strain $\gamma_{c}\equiv2\sqrt{E_{min}}$
(in our units%
\footnote{Both the shear transformation volume and the shear modulus are set
to unity, so that $E_y=\nicefrac{\gamma_y^2}{4}$.%
}) has been cumulated in the plastic region, i.e., when $\int\left\Vert 2\boldsymbol{\dot{\epsilon}}\left(t\right)\right\Vert dt=\gamma_{c}$.
Interestingly, this criterion, albeit rather arbitrary, captures the observed decrease of the rearrangement (T1 event) time at high enough strain rates
in bubble clusters \cite{Biance2009}.

\begin{figure}[htpb]
\begin{centering}
\includegraphics[width=\columnwidth]{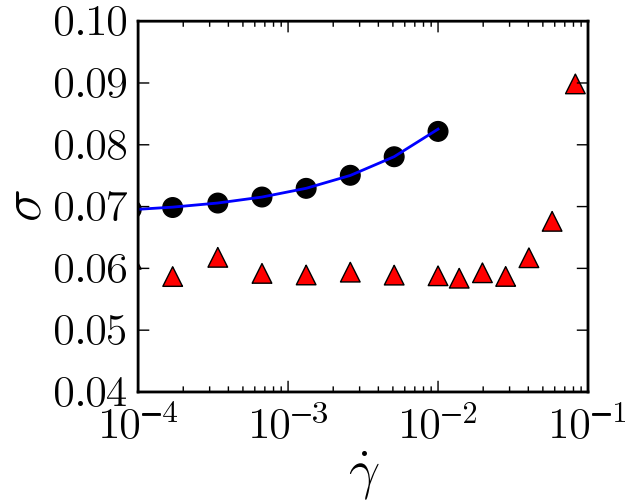}
\par\end{centering}

\caption{\label{fig:Flow-curves-refined}Flow curves $\text{\ensuremath{\sigma\left(\dot{\gamma}\right)}}$
for the convected tensorial model with refined probabilities and no
thermal activation ($x_{loc}=0$). The model parameters are: $\gamma_{c}=0.7$
and (\emph{triangles}) $k=10^{-3}$ or (\emph{dots}) $k=\infty$. The flow curve
for the infinite rate of recovery, $k=\infty$ is fit with a Herschel-Bulkley
equation, $\sigma=0.07+0.19\dot{\gamma}^{0.57}$.}

\end{figure}

Figure \ref{fig:Flow-curves-refined} presents the flow curve resulting
from these new dynamical rules, with $\gamma_{c}=0.7\left\langle \gamma_{y}\right\rangle $.
The curve is perfectly fit by a Herschel-Bulkley equation with exponent
$n\simeq0.57$ over a reasonable shear rate window. (At higher shear
rates, plastic events invade the whole simulation cell, and the dissipative
processes during plasticity dominate).

\subsection{Thermal activation of plastic events and ageing}

Thermally activated plastic events can easily be incorporated into
the model. This is achieved by modifying the yielding rate as follows,
\begin{equation}
l(\sigma,t)=\exp\left(\frac{E\left(\sigma\right)-E_{y}}{k_{B}T}\right),
\end{equation}
where $\sigma$ is the local magnitude of the shear stress, $E\left(\sigma\right)\equiv\nicefrac{\sigma^{2}}{4}$,
and $k_{B}T$ is the thermal energy.

In that case, thermally activated rejuvenation of the elastoplastic
blocks competes with the possibility to land in a deeper energy basin,
and Bouchaud's trap model\cite{Bouchaud1992} establishes the existence
of a liquid-glass transition at a critical temperature. Below this
temperature, the system spends most of the time exploring always deeper
wells on average after successive rearrangements, and the prevalence
of rearrangement-induced ageing of the material results in a macroscopic
yield stress as the shear rate goes to zero.

\subsection{Ageing and shear-banding}

The previous type of ageing requires particles to break out of the
cages created by their neighbours (only to end up in an even more
ridig cage). In a PEL perspective, this would correspond to successive
jumps between metabasins. Yet, one may think that, in some situations,
the energy minimisation within one metabasin, via jumps between basins,
can be only partially completed as the plastic event terminates, so
that short-term ageing can take place in the elastic regime. Physically,
candidates for short-term ageing would hypothetically be the reorientation
of particles in a Laponite suspension to maximise the electrostatic
or van der Waals interactions or the ion exchanges with the solvents.
For foams, it might be the time of recovery of the optimal angles
between bubble walls at the Plateau borders, or the equilibration time
of the surface tension of the liquid films upon a sudden variation of their area, in the case of foams with 
high surface modulus surfactants \cite{Denkov2009}. 
In granular matter, moisture-induced
ageing has been reported in the material at rest\cite{Bocquet1998}.
In our approach, short-term ageing shall straightforwardly be modelled
by imposing a finite time for the energy barrier $E_{y}\left(t\right)$
to reach its final value, after a plastic event, as follows,
\begin{equation}
\dot{E}_{y}\left(t\right)=k\frac{E_{y}-E_{y}\left(t\right)}{E_{y}-E_{min}},
\end{equation}
where $k$ is a rate of recovery, and stability is supposed to be
minimal at the end of a plastic event, that is, $E_{y}\left(t_{end}\right)=E_{min}$.
Note that the shear-induced lowering of the energy barriers has received
at least numerical confirmation \cite{Rodney2011}.

\begin{figure}[htpb]
\begin{centering}
\includegraphics[width=\columnwidth]{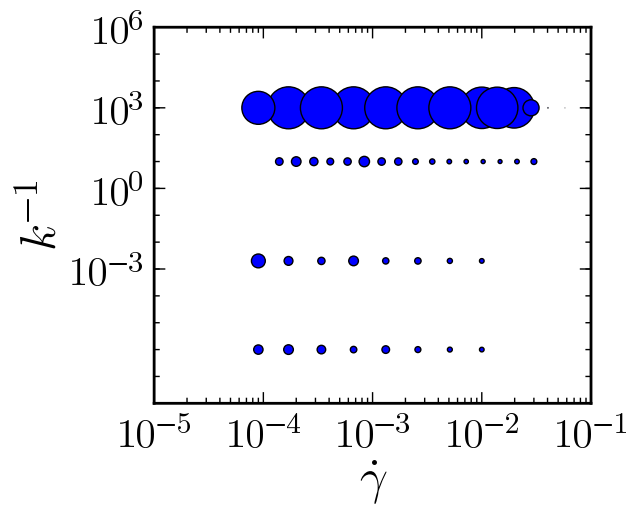}
\par\end{centering}

\caption{\label{fig:shear_banding_refined}Dependence of the shear-banding
parameter $\kappa$ on the applied shear rate $\dot{\gamma}$ and
the recovery time $k^{-1}$. The largest circle corresponds to $\kappa=1$.
The system consists of $128\times128$ blocks.}
\end{figure}

Qualitatively, the rate of recovery $k$ is analogous to the inverse of the restructuring time $\tau_{res}$ introduced in the simple model of section  \ref{sec:Simplistic-model}.
When $k$ is too low, and the driving
$\dot{\gamma}$ competes with the recovery process, the material will
be prone to shear localisation: fracture in a region  makes it durably
weaker. In Fig.\ref{fig:shear_banding_refined}, the extent of shear
localisation is quantified with the help of the shear-banding observable
$\kappa$ (see Section \ref{sub:Flow-curve-and-res-time}). The associated
stress plateau on the flow curve is conspicuous in  Fig. \ref{fig:Flow-curves-refined}.
At extremely low shear rates, $\dot{\gamma}\ll k$, one expects to
recover a homogeneous flow in the steady state, similar to that for
$k=\infty$ \cite{Chikkadi2012}. It is worth noting that Vandembroucq
and co-workers\cite{Vandembroucq2011} too have studied a coarse-grained
model with permanent strain weakening, which boils down to $k=0$
with our notations, and have also reported that it was associated
with shear localisation.

More generally, there is now growing evidence that the longer the
material needs to heal back to its pristine state after a plastic
event, the more prone it is to shear localisation. The general concept
of healing time can take diverse forms in practice: it can be the
duration of the rearrangement itself as in Ref.\cite{Coussot2010,Martens2012},
the chemical relaxation time of the local mechanical (e.g., surface tension) properties towards their equilibrium values
after a plastic event (which the authors of Ref.~\cite{Denkov2009} argue 
results in a flatter flow curve for foams with high surface modulus surfactants, 
as compared to their low surface modulus counterparts),
the thermally activated rebinding of failed contacts\cite{Kun2008},
or the time to dissipate the induced heat\cite{Shimizu2006}. Alternatively,
in its initial state, the material may have been stabilised by ageing
or a careful preparation protocol involving a slow cooling rate, the
benefits of which are lost locally upon the occurrence of a plastic
event\cite{Shimizu2006}. In this vein, Kumar et al.\cite{Kumar2013}
recently reported that the lower the fictive temperature of a metallic
glass, the more brittle it is. Also, attractive interactions are often
reported to enhance the heterogeneity of the flow, which
may be connected to long time scales to form stable aggregates\cite{Irani2013}. But the effect
of the increase transcends the variety of these mechanisms.

\subsection{4-point susceptibility $\chi_{4}$\label{sub:4-point-susceptibility-refined}}

Let us now set $k=\infty$ and investigate the four-point correlations
of the stress fluctuations, as in the previous model. The spatial
profiles of $\mathcal{G}_{4}\left(r,\Delta t^{\star}\right)$ (\emph{not
shown}) retain the symmetry described above, although in the convected
version the crosswise lobe is now less skewed, owing to the lower
yield strain. However, the integrals of $\mathcal{G}_{4}\left(r,\Delta t^{\star}\right)$
over growing disks centered at the origin scale differently with the
disk radii $R$; here they scale with $R^{\alpha}$ for $1\ll R\ll L$,
with $\alpha\approx0.6$ roughly in the static case.

\begin{figure}[htpb]
\begin{centering}
\subfloat[$\left(\alpha=1.5,\beta=0.94\right)$]{\begin{centering}
\includegraphics[width=0.5\columnwidth]{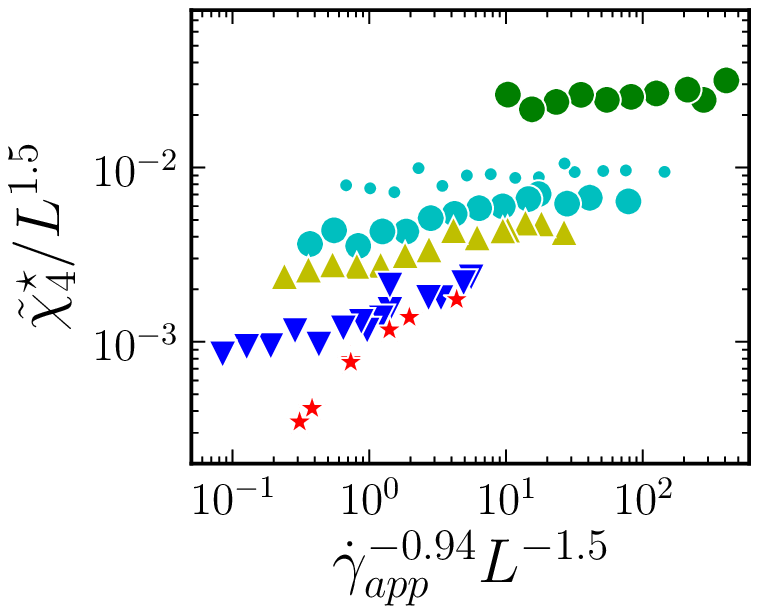}
\par\end{centering}

}\subfloat[$\left(\alpha=0.65,\beta=0.25\right)$]{\begin{centering}
\includegraphics[width=0.5\columnwidth]{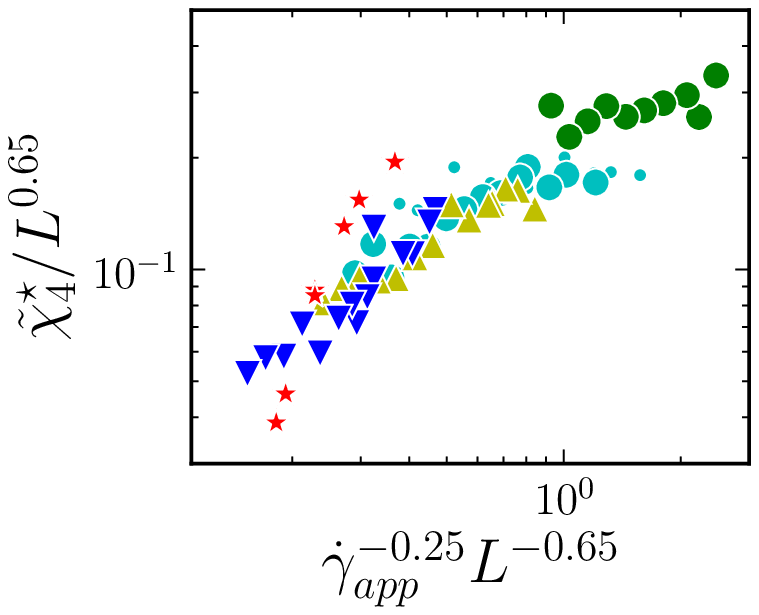}
\par\end{centering}

}
\par\end{centering}

\caption{\label{fig:chi4_NCBP-refined}Maximal cooperative volume $\tilde{\chi}_{4}^{\star}$,
rescaled according to $\frac{\tilde{\chi}_{4}}{L^{\alpha}}=f\left(\frac{\dot{\gamma}^{-\beta}}{L^{\alpha}}\right)$,
for two distinct pairs $\left(\alpha,\beta\right)$: the exponents
measured in the previous section, $\left(\alpha=1.5,\beta=0.94\right)$,
and better suited exponents, $\left(\alpha=0.65,\beta=0.25\right)$.
Various linear sizes of the (square) system are studied: $L=$ (\emph{large green
dots})16, (\emph{cyan dots}) $32$, (\emph{large cyan dots}) $48$, (\emph{yellow} \emph{triangles})
$64$, (\emph{blue} \emph{triangles}) $128$, (\emph{red} \emph{stars})
$256$. As a guide to the eye, we have plotted a dashed line with
slope 1. Data have been averaged over $\Delta\gamma\approx400\left\langle \gamma_{y}\right\rangle .$}
\end{figure}

To carry on with the quantitative study, we turn to the maximal cooperative
volume $\tilde{\chi}_{4}^{\star}$. Since no obvious scaling was found
in the convected case with the previous model, only the static tensorial
model is studied here. For a given system size, say, $L=128$, at
relatively high shear rates, the exponent $\beta$ in $\tilde{\chi}_{4}^{\star}\sim\dot{\gamma}^{-\beta}$
is of order 0.2-0.3, that is, significantly lower than its counterpart
for the simplistic model. In spite of the scatter of the data, we
clearly see that the scaling law used above is no longer valid for
this model, as illustrated in Fig.\ref{fig:chi4_NCBP-refined}; universality
in the variations of the cooperative volume with the system size and
the shear rate is thus ruled out. In fact, the scaling form $\frac{\tilde{\chi}_{4}}{L^{\alpha}}=f\left(\frac{\dot{\gamma}^{-\beta}}{L^{\alpha}}\right)$
in general does not seem to provide any nice collapse of the data
here.

\subsection{Cooperative disk}

As for the previous model, we determine, at an arbitrary point $M$
in the system, the radius $\xi^{\star}$ of the cooperative disk outside
which plastic events contribute to the instantaneous mechanical noise
perceived at $M$ essentially in a mean-field manner.

\begin{figure}[htpb]
\begin{centering}
\includegraphics[width=\columnwidth]{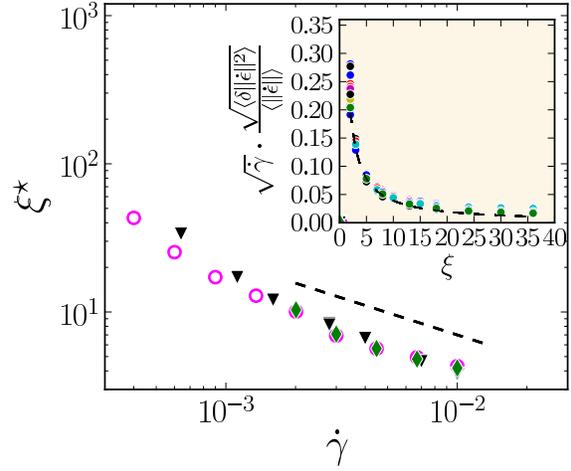}
\par\end{centering}

\caption{\label{fig:Cooperative-length-refined}Cooperative length $\xi^{\star}$
as a function of the applied shear rate $\dot{\gamma}$. Model parameters:
$\gamma_{c}=0.07$, $k=\infty,$ with linear system size $L=$(\emph{black
triangles}) 128, (\emph{pink circles}) 256, (\emph{green diamonds})
384. The dashed line has slope -0.5. (\emph{Inset}) Rescaled fluctuation
over mean value ratio $\sqrt{\dot{\gamma}}\frac{\sqrt{\left\langle \left\Vert \boldsymbol{\delta\dot{\epsilon}}\right\Vert {}^{2}\right\rangle }}{\left\Vert \left\langle \boldsymbol{\dot{\epsilon}}\right\rangle \right\Vert }$
as a function of $\xi$, for various $\dot{\gamma}$. The dashed line
represents $\nicefrac{0.39}{\xi}$.}

\end{figure}

The inset of Figure \ref{fig:Cooperative-length-refined} shows that
the scaling $\frac{\sqrt{\left\langle \left\Vert \boldsymbol{\delta\dot{\epsilon}}\right\Vert {}^{2}\right\rangle }}{\left\Vert \left\langle \boldsymbol{\dot{\epsilon}}\right\rangle \right\Vert }\left(\xi\right)\sim\frac{1}{\xi\sqrt{\dot{\gamma}}}$
is also entirely satisfactory in this case. It follows that the scaling
of the cooperative length $\xi^{\star}$ with $\dot{\gamma}^{-\nicefrac{1}{2}}$
is conserved, although a departure from this scaling is observed when
$\xi^{\star}\gtrsim20$, due to finite-size effects (see Fig. \ref{fig:Cooperative-length-refined}).
This supports the idea that the arguments developed above to ground
the scaling of this cooperative length with the spacing between simultaneous
plastic events are not model-specific; we expect them to be relevant,
for $\xi^{\star}$, even in the thermal regime.

\subsection{Persistent limits of the model, importance of fluctuations, outlook}

The refined model presented in this section features an additional
source of disorder as compared to the model with simplistic rules
presented in Section \ref{sec:Simplistic-model} and studied in Section
\ref{sec:Correlation-lengths}: a distribution of yield stresses.
Nevertheless, cooperative effects persist, and fluctuations of the
mechanical noise remain large, 
which may explain why deviations from
mean-field behaviours have been reported in the flow of amorphous
solids\cite{Lin2014,Budrikis2013}. 

It must however be said that some other aspects of disorder in these
systems are still ignored in our model; accordingly, the spatial correlations
that we predict with its help may well be overestimated. Among these
other aspects of disorder, let us mention the spatial heterogeneities
in shear modulus and the particulate nature of the material at the
scale of a shear transformation zone, both of which are expected to
alter the elastic propagator that we use. Shear waves are also expected
to be damped by the finite viscosity of the (visco)elastic medium,
but also by plastic regions, whose softening has been discarded so
far in the computation of the elastic propagator. Finally, we have
assumed an instantaneous propagation of shear waves.

\section{Conclusion}

In this contribution, we have presented a detailed study of diverse
variants of a coarse-grained model for the flow of amorphous solids.
These models consist of elastic blocks that yield when the local stress
gets too large; interaction between the blocks result from the stress
redistribution that takes place during the plastic events. In a first
version of the model, we have assessed the importance to use a tensorial
stress, instead of settling with its component along the macroscopic
shear, and to account for the convection of the blocks (in a coarse-grained
fashion). In the absence of convection, scalar and tensorial models
are extremely similar in terms of flow curve, tendency to shear localisation,
and magnitude of spatial correlations. The convected model mainly
differs from the static ones because of the restored asymmetry between
the flow and velocity-gradient directions: for instance, shear bands
can only be aligned along the flow direction, in agreement with experimental
observations. The convected model also features enhanced fluctuations,
and does not follow the same scaling for the four-point stress susceptibility.
In that regard, the influence of the practical way in which convection
is implemented in the model may however be determinant.

We have also proposed new dynamical rules to make the model more directly
related to real systems. Flow curves more similar to those typically
observed in experiments have been obtained; in addition, we have been
able to propose a somewhat more general interpretation of the tendency
to shear localise as the healing time following a rearrangement increases.

Turning to the correlation lengths, the origin of the decreasing cooperativity
as the shear rate increases is that plastic events screen each other.
To study the decrease quantitatively, we have defined a cooperative
length delimiting a region where the fluctuations due to individual
plastic events matter from an outer region which mainly acts in a
mean-field way. Along with other prescriptions, this definition gives
a correlation length that scales with the spacing between simultaneous
plastic events (as assessed to leading order, i.e., for homogeneously
distributed plastic events), that is, $\dot{\gamma}^{\nicefrac{-1}{2}}$
in 2D in the athermal regime; this scaling is robust to model variations.
On the other hand, the variations with the system size and the shear
rate of the cooperative volume assessed via the four-point stress
susceptibility seem to depend largely upon the model that is considered.

One is thus led to the conclusion that, notwithstanding the existence
of a major class of correlation lengths that scale with $\dot{\gamma}^{\nicefrac{-1}{d}}$
in \emph{d} dimensions, there exists no universal scaling behaviour
which would hold for all correlation length definitions and all systems.  The universality observed in the correlations of the mechanical noise field, and expressed by the radius of the cooperative disk $\xi^\star$, does not extend to more sophisticated observables that would for example quantify the shape and size of collective plastic events.



\section*{Acknowledgements}
We thank Luca Marradi for help with the computer cluster. J.-L.B. is supported by Institut Universitaire de France and by Grant No. ERC-2011-ADG20110209.
Most of the computations presented in this paper were performed using the CIMENT infrastructure (https://ciment.ujf-grenoble.fr), which is supported by the Rh$\hat{o}$ne-Alpes region (GRANT CPER07\_13 CIRA: http://ci-ra.org).



\providecommand*{\mcitethebibliography}{\thebibliography}
\csname @ifundefined\endcsname{endmcitethebibliography}
{\let\endmcitethebibliography\endthebibliography}{}

\appendix

\section{Derivation of the elastic propagator in the deformed frame\label{sec:Derivation_of_prop_in_deformed_frame}}

\begin{figure}[htpb]
\begin{centering}
\includegraphics[width=\columnwidth]{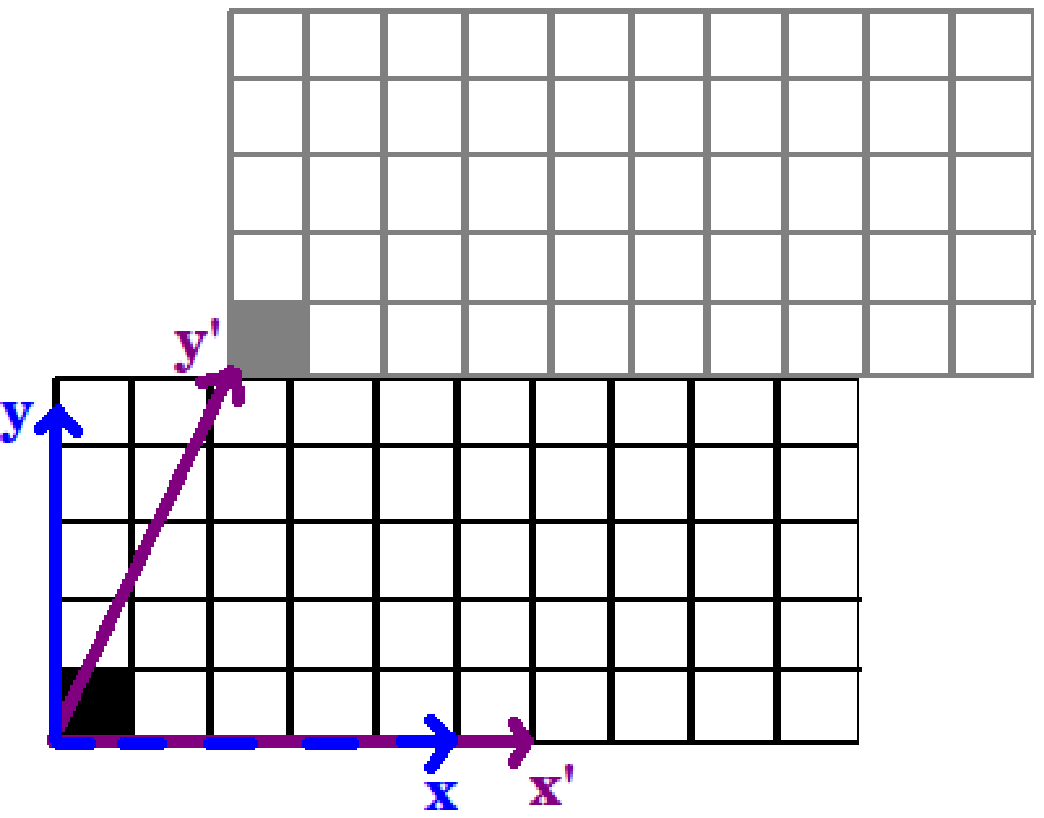}
\end{centering}
\par\end{figure}

In order to avoid a spurious discontinuity at the edge of the (periodic)
simulation cell when convection is included, the cell must to be deformed,
in a fashion equivalent to Lees-Edwards boundary conditions in atomistic
simulations: periodic replicas of the system in the velocity gradient-direction
shall be displaced along the flow, while replicas in the flow direction
remain unaffected. At a given time, the elastic propagator shall then
be expressed in a deformed frame, with (contravariant) coordinates
{$\left(x^{\prime},y^{\prime}\right)=\left(x-\gamma y,y\right)$},
where $\gamma$ is the average shear strain experienced by the cell.
Here, unprimed quantities refer to the initial orthonormal frame.
The corresponding transformation of the (covariant) Fourier wavenumbers
reads $q^{\prime}\equiv\left(q_{x}^{\prime},q_{y}^{\prime}\right)=\left(q_x,q_y+\gamma q_x\right)$.

The metric tensor,
\begin{equation}
\left(g_{ij}\right)\equiv\frac{\partial s^{k}}{\partial s^{\prime i}}\frac{\partial s^{l}}{\partial s^{\prime j}}\delta_{kl}=\left(\begin{array}{cc}
1 & \gamma\\
\gamma & 1+\gamma^{2}
\end{array}\right),\label{eq:metric_tensor}
\end{equation}
where $s=q$ or $s=(x,y)$, is a convenient tool to compute distances in the deformed frame; it
relates covariant and contravariant quantities, $x_{j}^{\prime}=g_{ij}x^{\prime i}$
as well as $q_{j}^{\prime}=g_{ij}q^{\prime i}$, for $i,\,j \in \{x,y\}$. When $\gamma\neq0$,
the metric tensor $g_{ij}$ differs from identity, so that $q^\prime_{i}\neq q^{\prime\,i}$.
Nevertheless, provided that this difference is accepted, Eq.~\ref{eq:displacement_non_orth}
still holds, \emph{viz.} 
\begin{equation}
u^{\prime i}(q^\prime)=\frac{-2\mathrm{i}}{q^{\prime 4}}
\left[q^{\prime 2}q_{k}^{\prime}
\epsilon^{\prime \mathrm{(pl)}\, ki}(q^\prime)-q^{\prime i}q_{k}^{\prime}q_{l}^{\prime}\epsilon^{\prime \mathrm{(pl)}\, kl}(q^\prime)
\right],
\label{eq:app:start_eq}
\end{equation}
from which the strain tensor $\epsilon^{\prime ij}=\frac{q^{\prime\,i}u^{\prime j}+q^{\prime\,j}u^{\prime i}}{2}$
and the elastic (deviatoric) stress $\sigma{}^{\prime ij}=2\mu\left(\epsilon^{\prime ij}-\epsilon^{\mathrm{(pl)}\, ij}\right)$
readily follow. As in Eq.~\ref{eq:Guillemette2}, the dots indicating time derivatives have been dropped
in Eq.~\ref{eq:app:start_eq}. Finally, the \emph{components} of the stress tensor
(and not the \emph{coordinates} of the points at which it is evaluated)
are to be expressed in the original, orthogonal basis $\left(x,y\right)$
, as follows:
\begin{eqnarray}
\sigma^{ij}(q^\prime) & 
= & 
\frac{\partial x^{i}}{\partial x^{\prime r}}\frac{\partial x^{j}}{\partial x^{\prime s}}\sigma^{\prime rs}(q^\prime)\nonumber \\& 
= & 
\frac{2\mu}{q^{\prime4}}\Big[q^{\prime2}\left(\mathcal{A}_{u}\mathcal{P}^{j}\epsilon^{\mathrm{(pl)}\, ui}(q^\prime)+\mathcal{A}_{u}\mathcal{P}^{i}\epsilon^{\mathrm{(pl)}\, uj}(q^\prime)\right) \nonumber  \\&
- & 2 \mathcal{P}^{i}\mathcal{P}^{j}\mathcal{A}_{u}\mathcal{A}_{v}\epsilon^{\mathrm{(pl)}\, uv}(q^\prime)\Big] 
- 2\mu\epsilon^{\mathrm{(pl)}\, ij}(q^\prime). \label{eq:appendix_epsilon_ij_for_prop}
\end{eqnarray}

Note that we have used the shorthands $q^{\prime2}=q_{k}^{\prime}q^{\prime k}=q^2$,
$\mathcal{P}^{i}\equiv\frac{\partial x^{i}}{\partial x^{\prime r}}q^{r}$,
and $\mathcal{A}_{u}\equiv\frac{\partial x^{\prime k}}{\partial x^{u}}q_{k}=\left(q_x,q_y-\gamma q_x\right)$.

Explicit evaluation of Eq.~\ref{eq:appendix_epsilon_ij_for_prop} with
the metric tensor (\ref{eq:metric_tensor}) leads to our final result:

\begin{equation}
\left(\begin{array}{c}
\sigma{}^{xx}\\
\sigma{}^{xy}
\end{array}\right)(q^\prime) = \boldsymbol{\mathcal{G}}^{\infty}\cdot\left(\begin{array}{c}
\epsilon^{pl\, xx}\\
\epsilon^{pl\, xy}
\end{array}\right)(q^\prime),
\end{equation}
with
\begin{equation}
\mathcal{G}^{\infty}\equiv\frac{1}{q^{\prime 4}}
\left[\begin{array}{cc}
-(q_x^{\prime2}-q_y^{(\gamma)\,2})^{2} & \,\,\,-2q_x^{\prime}q_y^{(\gamma)}(q_x^{\prime2}-q_y^{(\gamma)\,2})\\
-2q_x^{\prime}q_y^{(\gamma)}(q_x^{\prime 2}-q_y^{(\gamma)\,2}) & -4q_x^{\prime 2}q_y^{(\gamma)\,2}
\end{array}\right]
\end{equation}
and $q_y^{(\gamma)} \equiv \left(q_y^{\prime}-\gamma q_x^{\prime}\right)$.

With biperiodic boundary conditions, this propagator results in periodic
images of the plastic events that are not aligned along the velocity
gradient direction, but tilted with an {}``angle'' $\gamma$. Besides the
(long-range) effect of these periodic images, the shape of the elastic
propagator in real space should in principle be insensitive to the
frame in which it is computed. However, we would like to indicate
that the discrete nature and the symmetries of the meshgrid that we use
introduce some near-field dependence of $\mathcal{G}^{\infty}$ on
$\gamma$, up to a distance of few meshes away from the origin of
the plastic event. It is therefore important to keep $\gamma$ within
a relatively narrow range, here, $[-\nicefrac{1}{2},\nicefrac{1}{2}]$,
which is achieved thanks to the periodicity of the system in the flow
direction.

\section{Calculation of the displacement of streamlines induced by plastic events\label{sec:Displacement}}

The deformation is not strictly affine in the system. Consequently, on top of the average deformation of the cell,
we need to compute the displacement of each streamline so as to
be able to shift it adequately with respect to its neighbours. The average (non-affine) displacement in the flow direction $x$ on streamline $y_0$ that is
induced by plastic events reads,
\begin{eqnarray*}
\left\langle u_{x}\right\rangle _{x}\left(y_{0}\right) & \equiv & L^{-1}\int u_{x}\left(x,y_{0}\right)dx
\\
 & 
 = 
 & 
 \sum_{q_y}u_x(q_x=0,q_y)e^{iq_yy_0}
 \\
 & 
 = 
 & \sum_{q_y}\frac{-2i}{q_y}\epsilon_{xy}^{\mathrm{(pl)}}\left(q_x=0,q_y\right)e^{iq_y y_0},
\end{eqnarray*}
where the sums run over all relevant wavenumbers $q_{y}=\nicefrac{2\pi n}{L}$
and, as is now usual, we have dropped the (1)-superscripts and the
hats denoting Fourier transforms. To obtain the last equality, we
have made use of Eq.~\ref{eq:displacement_non_orth}. Finally, one arrives at,
\begin{eqnarray*}
\left\langle u_x \right\rangle _{x}(y_0) 
&
= 
&
\sum_{q_y}\frac{-2i}{q_y} e^{iq_y y_{0}} 
  \\
 &
 &
 \times \left[L^{-1}\sum_{y_{ev}}  \left\langle \epsilon_{xy}^{\mathrm{(pl)}}(x,y_{ev})\right\rangle _{x}
  e^{-iq_y y_{ev}}\right]
\\  
&
=
&
    \frac{-2i}{L}\sum_{y_{ev}}\left\langle \epsilon_{xy}^{\mathrm{(pl)}}\right\rangle _{x}\left(y_{ev}\right)\sum_{q_y}\frac{e^{iq_{y}\left(y_0-y_{ev}\right)}}{q_{y}}
\\
& 
=
& 
 \sum_{y_{ev}}\mathrm{sign}(y_{0}-y_{ev})\\
&
& 
 \times\left(1-\frac{2\left|y_{0}-y_{ev}\right|}{L}\right)\left\langle \epsilon_{xy}^{\mathrm{(pl)}}\right\rangle _{x}\left(y_{ev}\right),
\end{eqnarray*}
where the sum runs over all streamlines $y_{ev}$ and, to get the
last line, we have summed the second series over all $q_{y}=\nicefrac{2\pi n}{L},\ n\in\mathbb{Z}^{\star}$.

Whenever the cumulative displacement of a streamline in the flow direction
reaches the size of a block, it is shifted. As a technical detail,
note that we also regularly add a random displacement offset to \emph{all}
lines in order to prevent the spurious pinning in the simulation cell
of the streamlines that have zero average velocity. Otherwise, this spurious pinning 
might have jeopardised the translational invariance of the system along the velocity gradient direction.

\section{Estimation of the cooperative length $\xi^{\star}$\label{sec:Estimation-of-xi_star}}

Simple arguments based on the crude assumption of randomly distributed
plastic events explain the importance of the mechanical noise fluctuations
measured at an arbitrary point, say the origin $M$, and the scaling
law in $d=2$ dimensions, Eq.~\ref{eq:xi_star_scaling}, which is recalled
here:
\begin{equation}
\frac{\sqrt{\left\langle \left\Vert \boldsymbol{\delta\dot{\epsilon}}\right\Vert {}^{2}\right\rangle }}{\left\Vert \left\langle \boldsymbol{\dot{\epsilon}}\right\rangle \right\Vert }\left(\xi\right)\sim\frac{1}{\xi\sqrt{\dot{\gamma}}}.
\end{equation}

Denoting by $p\in[0,1]$ the average surface fraction covered by plastic events at
a given shear rate and calling $\dot{\epsilon}^{pl}\sim\frac{\gamma_{y}}{\tau}$
the typical plastic strain rate, the mean value of the mechanical
noise due to plastic events occurring farther than $\xi$ is:
\begin{eqnarray*}
\left\Vert \left\langle \boldsymbol{\dot{\epsilon}}\right\rangle \right\Vert  & \approx & \int_{\xi}^{L}\int_{\mathcal{S}^{d}}\frac{p\dot{\epsilon}^{pl}}{L^{d}}r^{d-1}drd\theta\\
 & \sim & \frac{p\dot{\epsilon}^{pl}\left(L^{d}-\xi^{d}\right)}{L^{d}}.
\end{eqnarray*}

Here, $\mathcal{S}^{d}$ denotes the unit sphere in \emph{d} dimensions,
and we have used that the spatially averaged contribution of a plastic
event to the stress field is of order $\frac{\epsilon^{pl}}{L^{d}}$.
Numerical prefactors are omitted.

Let us now turn to the fluctuations and start by computing $\delta\dot{\epsilon}\left(r\right)$,
the contribution of plastic events taking place in a shell $[r,r+1]$
centred at $M$ :
\begin{eqnarray*}
\delta\dot{\epsilon}\left(r\right)^{2} & = & \left(\int_{r}^{r+1}dr^{\prime}\int_{\mathcal{S}^{d}}r^{\prime d-1}d\theta\mathcal{G}\left(r^{\prime},\theta\right)n\left(r^{\prime},\theta\right)\dot{\epsilon}^{pl}\right)^{2}\\
 & \sim & 0+\int_{r}^{r+1}dr^{\prime}\int_{\mathcal{S}^{d}}r^{\prime d-1}d\theta\left(\frac{\cos\left(4\theta\right)}{r^{d}}n\left(r,\theta\right)\dot{\epsilon}^{pl}\right)^{2}\\
 & \sim & \frac{p\cdot\left(\dot{\epsilon}^{pl}\right)^{2}}{r^{d+1}},
\end{eqnarray*}

where $n\left(r^{\prime},\theta\right)=0$ or 1 is the plastic activity
at point $\left(r^{\prime},\theta\right)$, and, crudely, we have
assumed the absence of static spatial correlations between plastic
events outside a fully correlated unit volume $r^{\prime d-1}dr^{\prime}d\theta=1$,
i.e., the volume of a single plastic event.

The sum $\delta\dot{\epsilon}$ of the contributions of the concentric
shells of radius $r\geqslant\xi$ then reads:
\begin{eqnarray*}
\delta\dot{\epsilon}{}^{2} & \sim & \left[\sum_{r=\xi}^{L}\delta\dot{\epsilon}\left(r\right)\right]^{2}\\
 & \sim & \sum_{r=\xi}^{L}\delta\dot{\epsilon}\left(r\right)^{2}\\
 & \sim & p\cdot\left(\dot{\epsilon}^{pl}\right)^{2}\left[\frac{L^{d}-\xi^{d}}{L^{d}\xi^{d}}\right].
\end{eqnarray*}

We are now able to compute the fluctuations-to-average ratio in the limit
$\xi\ll L$, 
\begin{equation}
\frac{\sqrt{\delta\dot{\epsilon}{}^{2}}}{\left\Vert \left\langle \boldsymbol{\dot{\epsilon}}\right\rangle \right\Vert }\sim\frac{1}{\xi^{\nicefrac{d}{2}}\sqrt{p}}.
\end{equation}

To conclude, one just has to recall that, when thermal fluctuations
are negligible, the density of plastic events is proportional to $\dot{\gamma}$,
because the typical stress release per plastic event $\epsilon^{pl}$
shows no significant dependence on the applied shear rate.
\end{document}